\documentclass{amsart}%
\usepackage{amsfonts}
\usepackage{amsmath}
\usepackage{amssymb}
\usepackage{graphicx}%
\newtheorem{theorem}{Theorem}

\newtheorem{definition}[theorem]{Definition}

\newtheorem{proposition}[theorem]{Proposition}

\begin{document}

\title[Segal--Bargmann transform for noncompact symmetric spaces]{The Segal--Bargmann transform for noncompact symmetric spaces of the complex type}
\author{Brian C. Hall}
\address{University of Notre Dame\\
www.nd.edu/\~ \thinspace bhall}
\email{bhall{@}nd.edu}
\author{Jeffrey J. Mitchell}
\address{Robert Morris University}
\email{mitchellj{@}rmu.edu}
\date{January 2005}

\begin{abstract}
We consider the generalized Segal--Bargmann transform, defined in terms of the
heat operator, for a noncompact symmetric space of the complex type. For
radial functions, we show that the Segal--Bargmann transform is a unitary map
onto a certain $L^{2}$ space of meromorphic functions. For general functions,
we give an inversion formula for the Segal--Bargmann transform, involving
integration against an \textquotedblleft unwrapped\textquotedblright\ version
of the heat kernel for the dual compact symmetric space. Both results involve
delicate cancellations of singularities.

\end{abstract}
\maketitle
\tableofcontents

\section{Introduction\label{intro.sec}}

The Segal--Bargmann transform for $\mathbb{R}^{d}$ \cite{Se1, Se2, Se3, Ba1}
is a widely used tool in mathematical physics and harmonic analysis. The
transform is unitary map $C_{t}$ from $L^{2}(\mathbb{R}^{d})$ onto
$\mathcal{H}L^{2}(\mathbb{C}^{d},\nu_{t}),$ where $\nu_{t}$ is a certain
Gaussian measure on $\mathbb{C}^{d}$ (depending on a positive parameter $t$)
and where $\mathcal{H}L^{2}$ denotes the space of holomorphic functions that
are square integrable with respect to the indicated measure. (See Section
\ref{rn.sec} for details.) From the point of view of harmonic analysis, one
can think of the Segal--Bargmann transform as combining information about a
function $f(x)$ on $\mathbb{R}^{d}$ with information about the Fourier
transform $\hat{f}(\xi)$ into a single holomorphic function $(C_{t}%
f)(x+i\xi).$ From the point of view of quantum mechanics for a particle moving
in $\mathbb{R}^{d},$ one can think of the Segal--Bargmann transform as a
unitary map between the \textquotedblleft position Hilbert
space\textquotedblright\ $L^{2}(\mathbb{R}^{d})$ and the \textquotedblleft
phase space Hilbert space\textquotedblright\ $\mathcal{H}L^{2}(\mathbb{C}%
^{d},\nu_{t}).$ In this setting, the parameter $t$ can be interpreted as
Planck's constant. Conceptually, the advantage of applying the Segal--Bargmann
transform is that it gives a description of the state of the particle that is
closer to the underlying classical mechanics, because we now have a function
on the classical phase space rather than on the classical configuration space.
See Section 2, \cite{mexnotes}, and \cite{Fo} for more information about the
Segal--Bargmann transform for $\mathbb{R}^{d}$ and its uses.

In the paper \cite{H1}, the first author introduced a generalization of the
Segal--Bargmann transform in which the configuration space $\mathbb{R}^{d}$ is
replaced by a connected compact Lie group $K$ and the phase space
$\mathbb{C}^{d}$ is replaced by the complexification $K_{\mathbb{C}}$ of $K.$
(See also the expository papers \cite{bull,range,mexnotes}.) The complex group
$K_{\mathbb{C}}$ can also be identified in a natural way with the cotangent
bundle $T^{\ast}(K),$ which is the usual phase space associated to the
configuration space $K.$ A main result of \cite{H1} is a unitary map $C_{t}$
from $L^{2}(K)$ onto $\mathcal{H}L^{2}(K_{\mathbb{C}},\nu_{t})$, where
$\nu_{t}$ is a certain heat kernel measure on the complex group $K_{\mathbb{C}%
}.$ The transform itself is given by applying the time-$t$ heat operator to a
function $f$ in $L^{2}(K)$ and then analytically continuing the result from
$K$ to $K_{\mathbb{C}}.$ The paper \cite{H2} then gave an inversion formula
for $C_{t}$ in which to recover the function $f$ on $K$ one integrates the
holomorphic function $C_{t}f$ over each fiber in $T^{\ast}(K)\cong
K_{\mathbb{C}}$ with respect to a suitable heat kernel measure. See also
\cite{KTX} for a study of the Segal--Bargmann transform, defined in terms of
the heat operator, on the Heisenberg group.

The motivation for the generalized Segal--Bargmann transform for $K$ was work
of Gross in stochastic analysis, specifically the Gross ergodicity theorem
\cite{Gr} for the loop group over $K.$ See \cite{bull, GM, HS, ergodic} for
connections between the generalized Segal--Bargmann transform and stochastic
analysis. The generalized Segal--Bargmann transform has also been used in the
theory of loop quantum gravity \cite{A, Th, TW1, TW2, Das1, Das2}. It has a
close connection to the canonical quantization of $(1+1)$-dimensional
Yang--Mills theory \cite{Wr, DH1, ymcoherent}. It can be understood from the
point of view of geometric quantization \cite{geoquant, FMMN, FMMN2}. Most
recently, it has been used in studying nonabelian theta functions and the
conformal blocks in WZW conformal field theory \cite{FMN1, FMN2}. (See also
\cite{Ty}.) See the paper \cite{bull} for a survey of the generalized
Segal--Bargmann transform and related notions.

In the paper \cite{St}, Stenzel extended the results of \cite{H1, H2} from the
case of compact Lie groups to the case of general compact symmetric spaces. We
give here a schematic description of Stenzel's results; see Section
\ref{compact.sec} for details. If $X$ is a compact symmetric space, there is a
natural \textquotedblleft complexification\textquotedblright\ $X_{\mathbb{C}}$
of $X.$ There is a natural diffeomorphism between the cotangent bundle
$T^{\ast}(X)$ and the complexification $X_{\mathbb{C}}.$ Under this
diffeomorphism, each fiber in $T^{\ast}(X)$ maps to a set inside
$X_{\mathbb{C}}$ that can be identified with the \textit{dual noncompact
symmetric space to }$X.$ (For example, if $X$ is the $d$-sphere $S^{d},$ then
each fiber in $T^{\ast}(S^{d})$ gets identified with hyperbolic $d$-space.)
Thus the complexified symmetric space $X_{\mathbb{C}}$ is something like a
product of the compact symmetric space $X$ and the dual noncompact symmetric
space. Since each fiber in $T^{\ast}(X)\cong X_{\mathbb{C}}$ is identified
with this noncompact symmetric space, we can put on each fiber the
\textit{heat kernel measure} for that noncompact symmetric space (based at the
origin in the fiber).

The Segal--Bargmann transform now consists of applying the time-$t$ heat
operator to a function in $L^{2}(X)$ and analytically continuing the resulting
function to $X_{\mathbb{C}}.$ The first main result is an inversion formula:
to recover a function from its Segal--Bargmann transform, one simply
integrates the Segal--Bargmann transform over each fiber in $T^{\ast}(X)\cong
X_{\mathbb{C}}$ with respect to the appropriate heat kernel measure. The
second main result is an isometry formula: the $L^{2}$ norm of the original
function can be computed by integrating the absolute-value squared of the
Segal--Bargmann transform, first over each fiber using the heat kernel measure
and then over the base with using the Riemannian volume measure. See Theorem
\ref{compact.thm} in Section \ref{compact.sec} for details. See Section 3.4 of
\cite{bull} for more information on the transform for general compact
symmetric spaces and \cite{range,KR1, KR2, HM1, HM2} for more on the special
case in which $X$ is a $d$-sphere.

Since we now have a Segal--Bargmann transform for the Euclidean symmetric
space $\mathbb{R}^{d}$ and for compact symmetric spaces, it is natural to
consider also the case of noncompact symmetric spaces. Indeed, since the
duality relationship between compact and noncompact symmetric spaces is a
symmetric one, it might seem at first glance as if one might be able to simply
reverse the roles of the compact and the noncompact spaces to obtain a
transform starting on a noncompact symmetric space. Unfortunately, further
consideration reveals significant difficulties with this idea. First, if $X$
is a noncompact symmetric space, then the fibers in $T^{\ast}(X)$ are not
compact and therefore cannot be identified with the compact dual to $X.$ (For
example, if $X$ is hyperbolic $d$-space, then the fibers in $T^{\ast}(X)$ are
diffeomorphic to $\mathbb{R}^{d}$ and not to $S^{d}.$) Second, if one applies
the time-$t$ heat operator to a function on a noncompact symmetric space $X$
and then tries to analytically continue, one encounters singularities that do
not occur in the compact case.

The present paper is a first step in overcoming these difficulties. (See the
end of this section for other recent work in this direction.) We consider
noncompact symmetric spaces of the \textquotedblleft complex\textquotedblright%
\ type, namely, those that can be described as $G/K,$ where $G$ is a connected
\textit{complex} semisimple group and $K$ is a maximal compact subgroup of
$G.$ (The simplest example is hyperbolic 3-space.) The complex case is nothing
but the noncompact dual of the compact group case. For noncompact symmetric
spaces of the complex type, we obtain two main results.

Our first main result is an isometry formula for the Segal--Bargmann transform
on the space of radial functions. We state this briefly here; see Section
\ref{isometry.sec} for details. Consider a function $f$ in $L^{2}(G/K)$ ($G$
complex) that is \textquotedblleft radial\textquotedblright\ in the symmetric
space sense, that is, invariant under the left action of $K$ on $G/K.$ Let
$F=e^{t\Delta/2}f$ and consider the map%
\begin{equation}
X\rightarrow F(e^{X}),\quad X\in\mathfrak{p}, \label{fex1}%
\end{equation}
where the Lie algebra $\mathfrak{g}$ of $G$ is decomposed in the usual way as
$\mathfrak{g}=\mathfrak{k}+\mathfrak{p}.$ We show that the map (\ref{fex1})
has a \textit{meromorphic} (but usually not holomorphic) extension from
$\mathfrak{p}$ to $\mathfrak{p}_{\mathbb{C}}:=\mathfrak{p}+i\mathfrak{p}$. The
main result of Section \ref{isometry.sec} is that there exist a constant $c$
and a holomorphic function $\delta$ on $\mathfrak{p}_{\mathbb{C}}$ such that
for all radial $f$ in $L^{2}(G/K)$ we have%
\begin{equation}
\int_{G/K}\left\vert f(x)\right\vert ^{2}dx=e^{ct}\int_{\mathfrak{p}%
_{\mathbb{C}}}\left\vert F(e^{X+iY})\right\vert ^{2}\left\vert \delta
(X+iY)\right\vert ^{2}\frac{e^{-\left\vert Y\right\vert ^{2}/t}}{(\pi
t)^{d/2}}dX~dY,\quad F=e^{t\Delta/2}f. \label{isometry.intro}%
\end{equation}

There is a \textquotedblleft cancellation of singularities\textquotedblright%
\ occurring here: although in most cases the function $F(e^{X+iY})$ is
singular at certain points, the singularities occur only at points where
$\delta(X+iY)$ is zero. Thus, the singularities in $F(e^{X+iY})$ are canceled
by the zeros in the density of the measure occurring on the right-hand side of
(\ref{isometry.intro}). Furthermore, by considering radial functions, we are
introducing a distinguished basepoint (the identity coset). Thus, in the
radial case, we are able to use the complexified tangent space at the
basepoint (namely, $\mathfrak{p}_{\mathbb{C}}$) as our \textquotedblleft
complexification\textquotedblright\ of $G/K,$ and we simply do not attempt to
identify $\mathfrak{p}_{\mathbb{C}}$ with $T^{\ast}(G/K).$ Of course, because
we are treating the identity coset differently from other points, this
approach is not $G$-invariant and is not the correct approach for the general
(nonradial) case.

Our second main result is an inversion formula for the Segal--Bargmann
transform of general (not necessarily radial) functions. We state this briefly
here; see Section \ref{inversion.sec} for details. We continue to assume that
$G$ is a connected \textit{complex} semisimple group and $K$ a maximal compact
subgroup. For each point $x$ in $G/K$, we have the geometric exponential map
$\exp_{x}$ taking the tangent space $T_{x}(G/K)$ into $G/K.$ Let $f$ be in
$L^{2}(G/K)$ and let $F=e^{t\Delta/2}f.$ Then, for each $x\in G/K,$ the
function%
\begin{equation}
X\rightarrow F(\exp_{x}X),\quad X\in T_{x}(G/K), \label{fex2}%
\end{equation}
admits an analytic continuation to some ball around zero. For each $x\in G/K,$
define%
\[
G(x,R)=e^{ct/2}\int_{\substack{Y\in T_{x}(G/K)\\\left\vert Y\right\vert \leq
R}}F(\exp_{x}iY)\delta(iY)\frac{e^{-\left\vert Y\right\vert ^{2}/2t}}{(2\pi
t)^{d/2}}~dY
\]
for all sufficiently small $R.$ (Here the constant $c$ and the function
$\delta$ are the same as in the isometry formula (\ref{isometry.intro}).)

Our main result is that for each $x$ in $G/K,$ $G(x,R)$ admits a real-analytic
continuation in $R$ to $(0,\infty)$ and, if $f$ is sufficiently regular,%
\[
f(x)=\lim_{R\rightarrow\infty}G(x,R).
\]
We may write this informally as%
\begin{equation}
f(x)=\text{ \textquotedblleft}\lim_{R\rightarrow\infty}%
\text{\textquotedblright\ }e^{ct/2}\int_{\left\vert Y\right\vert \leq R}%
F(\exp_{x}iY)\delta(iY)\frac{e^{-\left\vert Y\right\vert ^{2}/2t}}{(2\pi
t)^{d/2}}~dY, \label{inversion.intro}%
\end{equation}
where the expression \textquotedblleft$\lim_{R\rightarrow\infty}{}%
$\textquotedblright\ means that we interpret the right-hand side of
(\ref{inversion.intro}) literally for small $R$ and then extend to large $R$
by means of analytic continuation.

As in the isometry formula for radial functions, there is a cancellation of
singularities here that allows $G(x,R)$ to extend analytically to
$(0,\infty),$ even though $F(\exp_{x}iY)$ itself may have singularities for
large $Y.$ Because of the rotationally invariant nature of the integral in
(\ref{inversion.intro}), the integral only \textquotedblleft
sees\textquotedblright\ the part of the function $F(\exp_{x}iY)$ that is
rotationally invariant. Taking the rotationally invariant part eliminates some
of the singularities in $F(\exp_{x}iY).$ The remaining singularities are
canceled by the zeros in the function $\delta(iY).$

The measure against which we are integrating $F(\exp_{x}iY)$ in
(\ref{inversion.intro}), namely,
\[
d\sigma_{t}(Y)=e^{ct/2}\delta(iY)\frac{e^{-\left\vert Y\right\vert ^{2}/2t}%
}{(2\pi t)^{d/2}}~dY,
\]
is closely related to the heat kernel measure on the \textit{compact}
symmetric space dual to $G/K.$ Specifically, it is an \textquotedblleft
unwrapped\textquotedblright\ version of that heat kernel measure, in a precise
sense described in Section \ref{inversion.sec}.

The papers \cite{H2} and \cite{St} use the inversion formula for the
Segal--Bargmann transform (for compact groups and compact symmetric spaces,
respectively) to deduce the isometry formula. Since we now have an inversion
formula for the Segal--Bargmann transform for noncompact symmetric spaces of
the complex type, it is reasonable to hope to obtain an isometry formula as
well, following the line of reasoning in \cite{H2} and \cite{St}. The
hoped-for isometry formula in the complex case would involve integrating
$\left\vert F\right\vert ^{2}$ over a tube of radius $R$ (with respect to the
appropriate measure) and then analytically continuing with respect to $R.$
Since, however, there are many technicalities to attend to in carrying out
this idea, we defer this project to a future paper. (See \cite{range} for an
additional discussion of this matter.)

Meanwhile, it would be desirable to extend the results of this paper to other
symmetric spaces of the noncompact type. Unfortunately, the singularities that
occur in general are worse than in the complex case and are not as easily
canceled out. We discuss the prospects for other symmetric spaces in Section
\ref{conclude.sec}.

We conclude this introduction by comparing our work here to other types of
Segal--Bargmann transform for noncompact symmetric spaces. First, \'{O}lafsson
and \O rsted \cite{OO} have introduced another sort of Segal--Bargmann
transform for noncompact symmetric spaces, based on the \textquotedblleft
restriction principle.\textquotedblright\ This has been developed in
\cite{DOZ1,DOZ2} and used to study Laplace transforms and various classes of
orthogonal polynomials connected to noncompact symmetric spaces. This
transform does not involve the heat operator and is thus not directly
comparable to the Segal--Bargmann transform in this paper.

Meanwhile, Kr\"{o}tz, \'{O}lafsson, and Stanton \cite{KS1, KS2, KOS} have
considered the Segal--Bargmann transform for a general symmetric space $G/K$
of the noncompact type (not necessarily of the complex type), defined in the
same way as here, in terms of the heat equation. In \cite{KS2}, Kr\"{o}tz and
Stanton identify the maximal domain inside $G_{\mathbb{C}}/K_{\mathbb{C}}$ to
which a function of the form $e^{t\Delta/2}f$ can be analytically continued.
Then in \cite{KOS}, Kr\"{o}tz, \'{O}lafsson, and Stanton give an isometry
result identifying the image of $L^{2}(G/K)$ under the Segal--Bargmann
transform in terms of certain orbital integrals. There is also a cancellation
of singularities in their approach, in that the pseudodifferential operator
$D$ in Theorem 3.3 of \cite{KOS} is used to extend the orbital integrals into
the range where the function involved becomes singular. It remains to be
worked out how the results of \cite{KOS} relate, in the complex case, to the
isometry result suggested by the results we obtain in this paper.

\section{Review of the $\mathbb{R}^{d}$ case\label{rn.sec}}

We give here a very brief review of results concerning the Segal--Bargmann
transform for $\mathbb{R}^{d}.$ We do this partly to put into perspective the
results for noncompact symmetric spaces and partly because we will use the
$\mathbb{R}^{d}$ results in our analysis of the symmetric space case. See also
Section \ref{compact.sec} for a description of Stenzel's results for the case
of compact symmetric spaces.

In the $\mathbb{R}^{d}$ case, we consider the \textquotedblleft
invariant\textquotedblright\ form of the the Segal--Bargmann transform, which
uses slightly different normalization conventions from Segal \cite{Se3} or
Bargmann \cite{Ba1}. (See \cite{mexnotes} or \cite{newform} for a comparison
of normalizations.) The transform is the map $C_{t}$ from $L^{2}%
(\mathbb{R}^{d})$ into the space $\mathcal{H}(\mathbb{C}^{d})$ of holomorphic
functions on $\mathbb{C}^{d}$ given by%
\[
(C_{t}f)(z)=\int_{\mathbb{R}^{d}}(2\pi t)^{-d/2}e^{-(z-x)^{2}/2t}f(x)~dx,\quad
z\in\mathbb{C}^{d}.
\]
Here $(z-x)^{2}=(z_{1}-x_{1})^{2}+\cdots+(z_{d}-x_{d})^{2}$ and $t$ is an
arbitrary positive parameter. It is not hard to show that the integral is
convergent for all $z\in\mathbb{C}^{d}$ and the result is a holomorphic
function of $z.$

Recognizing that the function $(2\pi t)^{-d/2}e^{-(z-x)^{2}/2t}$ is (for $z$
in $\mathbb{R}^{d}$) the heat kernel for $\mathbb{R}^{d},$ we may also
describe $C_{t}f$ as%
\[
C_{t}f=\text{analytic continuation of }e^{t\Delta/2}f.
\]
Here the analytic continuation is from $\mathbb{R}^{d}$ to $\mathbb{C}^{d}$
with $t$ fixed. We take the Laplacian $\Delta=\Sigma\partial^{2}/\partial
x_{k}^{2}$ to be a negative operator, so that $e^{t\Delta/2}$ is the
\textit{forward} heat operator.

\begin{theorem}
[Segal--Bargmann]\label{rn.thm}Let $f$ be in $L^{2}(\mathbb{R}^{d})$ and let
$F=C_{t}f.$ Then we have the following results.

1. The \textbf{inversion formula}. If $f$ is sufficiently regular we have%
\begin{equation}
f(x)=\int_{\mathbb{R}^{d}}F(x+iy)\frac{e^{-y^{2}/2t}}{(2\pi t)^{d/2}}~dy
\label{rn.inv}%
\end{equation}
with absolute convergence of the integral for all $x.$

2. The \textbf{isometry formula}. For all $f$ in $L^{2}(\mathbb{R}^{d})$ we
have%
\begin{equation}
\int_{\mathbb{R}^{d}}\left\vert f(x)\right\vert ^{2}dx=\int_{\mathbb{R}^{d}%
}\int_{\mathbb{R}^{d}}\left\vert F(x+iy)\right\vert ^{2}\frac{e^{-y^{2}/t}%
}{(\pi t)^{d/2}}~dy~dx. \label{rn.isom}%
\end{equation}

3. The \textbf{surjectivity theorem}. For any holomorphic function $F$ on
$\mathbb{C}^{d}$ such that the integral on the right-hand side of
(\ref{rn.isom}) is finite, there exists a unique $f$ in $L^{2}$ with
$F=C_{t}f.$
\end{theorem}

The reason for the \textquotedblleft sufficiently regular\textquotedblright%
\ assumption in the inversion formula is to guarantee the convergence of the
integral on the right-hand side of (\ref{rn.inv}). It suffices to assume that
$f$ has $n$ derivatives in $L^{2}(\mathbb{R}^{d}),$ with $n>d/2.$ (See Section
2.1 of \cite{range}.)

The isometry and surjectivity formulas are obtained by adapting results of
Segal \cite{Se3} or Bargmann \cite{Ba1} to our normalization of the transform.
The inversion formula is elementary (e.g., \cite{range}) but does not seem to
be as well known as it should be. The inversion formula is implicit in Theorem
3 of \cite{Se0} and is essentially the same as the inversion formula for the
$S$-transform in \cite[Theorem 4.3]{Ku}. In quantum mechanical language, the
inversion formula says that the \textquotedblleft position wave
function\textquotedblright\ $f(x)$ can be obtained from the \textquotedblleft
phase space wave function\textquotedblright\ $F(x+iy)$ by integrating out the
momentum variables (with respect to a suitable measure).

It should be noted that because $F(x+iy)$ is holomorphic, there can be many
different inversion formulas, that is, many different integrals involving
$F(x+iy)$ all of which yield the value $f(x).$ For example, we may think of
the heat operator as a unitary map from $L^{2}(\mathbb{R}^{d})$ to the Hilbert
space of holomorphic functions for which the right-hand side of (\ref{rn.isom}%
) is finite. Then we may obtain one inversion formula by noting that the
adjoint of a unitary map is its inverse. The resulting \textquotedblleft
inverse = adjoint\textquotedblright\ formula is sometimes described as
\textquotedblleft the\textquotedblright\ inversion formula for the
Segal--Bargmann transform. Nevertheless, the inversion formula in
(\ref{rn.inv}) is \textit{not} the one obtained by this method.

In light of what we are going to prove in Section \ref{isometry.sec}, it is
worth pointing out that we could replace \textquotedblleft
holomorphic\textquotedblright\ with \textquotedblleft
meromorphic\textquotedblright\ in the statement of Theorem \ref{rn.thm}. That
is, we could describe $F$ as the meromorphic extension of $e^{t\Delta/2}f$
from $\mathbb{R}^{d}$ to $\mathbb{C}^{d}$ (if $F$ is holomorphic then it is
certainly meromorphic), and we could replace the surjectivity theorem by
saying that if $F$ is any meromorphic function for which the integral on the
right-hand side of (\ref{rn.isom}) is finite arises as the meromorphic
extension of $e^{t\Delta/2}f$ for some $f$ in $L^{2}(\mathbb{R}^{d}).$ After
all, since the density in (\ref{rn.isom}) is strictly positive everywhere,
such an $F$ would have to be locally square-integrable with respect to
Lebesgue measure, and it is not hard to show that a meromorphic function with
this property must actually be holomorphic. (This can be seen from the
Weierstrass Preparation Theorem \cite[p. 8]{GH}.) That is, under the
assumption that the right-hand side of (\ref{rn.isom}) is finite, meromorphic
and holomorphic are equivalent.

\section{Isometry for radial functions\label{isometry.sec}}

In this section we describe an isometric version of the Segal--Bargmann
transform for \textquotedblleft radial\textquotedblright\ functions on a
noncompact symmetric space $X$ of the \textquotedblleft complex
type\textquotedblright\ (e.g., hyperbolic 3-space).\ We give two different
forms of this result. The first involves integration over the complexified
tangent space to the symmetric space at the basepoint. The second involves
integration over the complexified tangent space to the maximal flat at the
base point. Both results characterize the image under the Segal--Bargmann
transform of the radial subspace of $L^{2}(X)$ as a certain holomorphic
$L^{2}$ space of \textit{meromorphic} functions. In Section \ref{conclude.sec}%
, we discuss the prospects for extending these results to nonradial function
and to other symmetric spaces of the noncompact type.

If $f$ is a function on a noncompact symmetric space $X=G/K$, then we wish to
define the Segal--Bargmann transform of $f$ to be some sort of analytic
continuation of the function $F:=e^{t\Delta/2}f.$ The challenge in the
noncompact case is to figure out precisely what sort of analytic continuation
is the right one. One could try to analytically continue to $G_{\mathbb{C}%
}/K_{\mathbb{C}},$ but examples show that $F$ does not in general admit an
analytic continuation to $G_{\mathbb{C}}/K_{\mathbb{C}}.$ Alternatively, one
could consider the maximal domain $\Omega$ to which functions of the form
$F=e^{t\Delta/2}f$ actually have an analytic continuation. This domain was
identified by Kr\"{o}tz and Stanton \cite[Thm. 6.1]{KS2} as the
Akhiezer--Gindikin \textquotedblleft crown domain\textquotedblright\ in
$G_{\mathbb{C}}/K_{\mathbb{C}}.$ Unfortunately, it seems that there can be no
measure $\mu$ on $\Omega$ such that the map sending $f$ to the analytic
continuation of $F$ is an isometry of $L^{2}(G/K)$ into $L^{2}(\Omega,\mu).$
(See the discussion in \cite[Remark 3.1]{KOS}.) Thus, to get an isometry
result of the sort that we have in the $\mathbb{R}^{d}$ case and the compact
case, we must venture beyond the domain $\Omega$ into the region where $F$ has
singularities and find a way to deal with those singularities.

In this section, we assume that the symmetric space is of the complex type and
that $f$ (and thus also $F$) is radial. We then write $F$ in exponential
coordinates at the basepoint, which makes $F$ a function on the tangent space
at the basepoint. We show that $F$ admits a \textit{meromorphic} extension to
the complexified tangent space at the basepoint. This meromorphic extension of
$F$ is then square-integrable with respect to a suitable measure; the zeros in
the density of the measure cancel the singularities in $F.$ We obtain in this
way an isometry of the radial part of $L^{2}(X)$ onto a certain $L^{2}$ space
of meromorphic functions.

In the next section, we consider the more complicated case of nonradial
functions. We obtain there an inversion formula involving a more subtle type
of cancellation of singularities.

The set-up is as follows. We let $G$ be a connected \textit{complex}
semisimple group and $K$ a maximal compact subgroup of $G.$ Since $G$ is
complex, $K$ will be a compact real form of $G$. We decompose $\mathfrak{g}$
as $\mathfrak{g}=\mathfrak{k}+\mathfrak{p},$ where $\mathfrak{p}%
=i\mathfrak{k}.$ We then choose an inner product on $\mathfrak{p}$ that is
invariant under the adjoint action of $K.$ We consider the manifold $G/K$ and
we think of the tangent space at the identity coset to $G/K$ as the space
$\mathfrak{p}.$ There is then a unique $G$-invariant Riemannian structure on
$G/K$ whose value at the identity is the given inner product on $\mathfrak{p}%
.$ Then $G/K$ is a Riemannian symmetric space of the \textquotedblleft complex
type.\textquotedblright\ 

We emphasize that the word \textquotedblleft complex\textquotedblright\ here
does \textit{not} mean that $G/K$ is a complex manifold but rather that $G$ is
a complex Lie group. The complex structure on $G$ will play no direct role in
any definitions or proofs; for example, we will never consider holomorphic
functions on $G.$ Nevertheless, the complex\ case is quite special among all
symmetric spaces of the noncompact type (i.e., compared to spaces of the form
$G/K$ with $G$ \textit{real} semisimple and $K$ maximal compact). What is
special about the complex case is not the complex structure \textit{per se},
but rather the structure of the root system for $G/K$ in this case: it is a
reduced root system in which all roots have multiplicity 2. Still, it is
easier to say \textquotedblleft complex\textquotedblright\ than to say
\textquotedblleft reduced root system with all roots having multiplicity
2\textquotedblright! The simplest example of a noncompact symmetric space of
the complex type is hyperbolic 3-space, and this is the only hyperbolic space
that is of the complex type.

We will make use of special intertwining formulas for the Laplacian that hold
only in the complex case. (See the proof of Theorem \ref{isometry.thm1} for a
discussion of why the intertwining formulas hold only in this case.)
Nevertheless, there is hope for obtaining similar but less explicit results
for other symmetric spaces of the noncompact type. See Section
\ref{conclude.sec} for a discussion.

We consider the geometric exponential mapping for $G/K$ at the identity coset.
This coincides with the group-theoretical exponential mapping in the sense
that if we identify the tangent space at the identity coset with
$\mathfrak{p},$ then the geometric exponential of $X\in\mathfrak{p}$ is just
the coset containing the exponential of $X$ in the Lie-group sense. In this
section, we will use the notation $e^{X}$ to denote the the geometric
exponential at the identity coset of a vector $X$ in $\mathfrak{p}.$ We let
$\delta$ be the square root of the Jacobian of the exponential mapping at the
identity coset. This is the positive function satisfying%
\begin{equation}
\int_{G/K}f(x)~dx=\int_{\mathfrak{p}}f(e^{X})\delta(X)^{2}dX,
\label{volume.relation}%
\end{equation}
where $dx$ is the Riemannian volume measure on $G/K$ and where $dX$ is the
Lebesgue measure on $\mathfrak{p}$ (normalized by the inner product).
Explicitly, $\delta$ is the unique Ad-$K$-invariant function on $\mathfrak{p}$
whose restriction to a maximal commutative subspace $\mathfrak{a}$ is given by%
\begin{equation}
\delta(H)=\prod_{\alpha\in R^{+}}\frac{\sinh\alpha(H)}{\alpha(H)}.
\label{delta.form}%
\end{equation}
Here $R$ is the set of (restricted) roots for $G/K$ (relative to
$\mathfrak{a}$) and $R^{+}$ is the set of positive roots relative to some
fixed Weyl chamber in $\mathfrak{a}.$ The expression (\ref{delta.form}) may be
obtained by specializing results \cite[Thm. IV.4.1]{He1} for general symmetric
spaces of the noncompact type to the complex case, in which \textit{all roots
have multiplicity two}. (Compare Equation (14) in Section V.5 of \cite{He3}.)

We consider functions on $G/K$ that are \textquotedblleft
radial\textquotedblright\ in the symmetric space sense, meaning invariant
under the left action of $K.$ (These functions are not necessarily functions
of the distance from the identity coset, except in the rank-one case.) We give
two isometry results, one involving integration over $\mathfrak{p}%
_{\mathbb{C}}:=\mathfrak{p}+i\mathfrak{p}$ and one involving integration over
$\mathfrak{a}_{\mathbb{C}}:=\mathfrak{a}+i\mathfrak{a}.$

\begin{theorem}
\label{isometry.thm1}Let $f$ be a radial function in $L^{2}(G/K)$ ($G$
complex) and let $F=e^{t\Delta_{G/K}/2}f.$ Then the function%
\begin{equation}
X\rightarrow F(e^{X}),\quad X\in\mathfrak{p}, \label{meromorphic1}%
\end{equation}
has a meromorphic extension from $\mathfrak{p}$ to $\mathfrak{p}_{\mathbb{C}}$
and this meromorphic extension satisfies%
\begin{equation}
\int_{G/K}\left\vert f(x)\right\vert ^{2}dx=e^{ct}\int_{\mathfrak{p}%
_{\mathbb{C}}}\left\vert F(e^{X+iY})\right\vert ^{2}\left\vert \delta
(X+iY)\right\vert ^{2}\frac{e^{-\left\vert Y\right\vert ^{2}/t}}{(\pi
t)^{d/2}}~dY~dX. \label{isometry1}%
\end{equation}
Here $c$ is the norm-squared of half the sum (with multiplicities) of the
positive roots for $G/K,$ and $d=\dim(G/K).$

Conversely, suppose $\Phi$ is a meromorphic function on $\mathfrak{p}%
_{\mathbb{C}}$ that is invariant under the adjoint action of $K$ and that
satisfies%
\begin{equation}
e^{ct}\int_{\mathfrak{p}_{\mathbb{C}}}\left\vert \Phi(X+iY)\right\vert
^{2}\left\vert \delta(X+iY)\right\vert ^{2}\frac{e^{-\left\vert Y\right\vert
^{2}/t}}{(\pi t)^{d/2}}~dY~dX<\infty. \label{finiteness1}%
\end{equation}
Then there exists a unique radial function $f$ in $L^{2}(G/K)$ such that
\[
\Phi(X)=(e^{t\Delta_{G/K}/2}f)(e^{X})
\]
for all $X\in\mathfrak{p}.$
\end{theorem}

On the right-hand side of (\ref{isometry1}), the expression $F(e^{X+iY})$
means the meromorphic extension of the function $X\rightarrow F(e^{X}),$
evaluated at the point $X+iY.$ The proof will show that $F(e^{X+iY}%
)\delta(X+iY)$ is holomorphic (not just meromorphic)\ on $\mathfrak{p}%
_{\mathbb{C}}.$ This means that although $F(e^{X+iY})$ will in most cases have
singularities, these singularities can be canceled out by multiplying by
$\delta(X+iY).$ This cancellation of singularities is the reason that the
integral on the right-hand side of (\ref{isometry1}) is even \textit{locally}
finite. Note that in contrast to the $\mathbb{R}^{d}$ case (where the density
of the relevant measure is nowhere zero), there exist here meromorphic
functions $F$ that are not holomorphic and yet are square-integrable with
respect to the measure in (\ref{isometry1}). Theorem \ref{isometry.thm1} holds
also for the Euclidean symmetric space $\mathbb{R}^{d},$ where in that case
$e^{X+iY}=X+iY,$ $c=0,$ and $\delta\equiv1,$ so that we have (\ref{rn.isom})
in the case where $f$ happens to be radial.

Observe that if $f$ is radial, then $F=e^{t\Delta/2}f$ is also radial. Thus
$F$ is determined by its values on a \textquotedblleft maximal
flat\textquotedblright\ $A:=\exp\mathfrak{a},$ where $\mathfrak{a}$ is any
fixed maximal commutative subspace of $\mathfrak{p}.$ Thus it is reasonable to
hope that we could replace the right-hand side of (\ref{isometry1}) with an
expression involving integration only over $\mathfrak{a}_{\mathbb{C}}.$ Our
next result is of this sort. We fix a Weyl chamber in $\mathfrak{a}$ and let
$R^{+}$ be the positive roots relative to this chamber. We let $\eta$ be the
function on $\mathfrak{a}$ given by%
\[
\eta(H)=\delta(H)\prod_{\alpha\in R^{+}}\alpha(H)=\prod_{\alpha\in R^{+}}%
\sinh\alpha(H).
\]
This function has an analytic continuation to $\mathfrak{a}_{\mathbb{C}},$
also denoted $\eta.$

\begin{theorem}
\label{isometry.thm2}Let $f$ be a radial function in $L^{2}(G/K)$ ($G$
complex) and let $F=e^{t\Delta_{G/K}/2}f.$ Then the function%
\[
H\rightarrow F(e^{H}),\quad H\in\mathfrak{a},
\]
has a meromorphic extension to $\mathfrak{a}_{\mathbb{C}}$ and this
meromorphic extension satisfies%
\begin{equation}
\int_{G/K}\left\vert f(x)\right\vert ^{2}dx=Be^{ct}\int_{\mathfrak{a}%
_{\mathbb{C}}}\left\vert F(e^{H+iY})\right\vert ^{2}\left\vert \eta
(H+iY)\right\vert ^{2}\frac{e^{-\left\vert Y\right\vert ^{2}/t}}{(\pi
t)^{r/2}}~dY~dH, \label{isometry2}%
\end{equation}
where $r=\dim\mathfrak{a}$ is the rank of $G/K$ and $c$ is as in Theorem
\ref{isometry.thm1}. Here $B$ is a constant independent of $f$ and $t.$

Conversely, suppose $\Phi$ is a meromorphic function on $\mathfrak{a}%
_{\mathbb{C}}$ that is invariant under the action of the Weyl group and that
satisfies%
\begin{equation}
Be^{ct}\int_{\mathfrak{a}_{\mathbb{C}}}\left\vert \Phi(H+iY)\right\vert
^{2}\left\vert \eta(H+iY)\right\vert ^{2}\frac{e^{-\left\vert Y\right\vert
^{2}/t}}{(\pi t)^{r/2}}\ dY~dH<\infty. \label{finiteness2}%
\end{equation}
Then there exists a unique radial function $f$ in $L^{2}(G/K)$ such that%
\[
\Phi(H)=(e^{t\Delta_{G/K}/2}f)(e^{H})
\]
for all $H\in\mathfrak{a}.$
\end{theorem}

In the dual compact case, an analogous result was established by Florentino,
Mour\~{a}o, and Nunes \cite[Thm. 2.2]{FMN2} and is described in Theorem
\ref{compact.radial} in Section \ref{compact.sec}.

Note that the function $F(e^{X+iY})$ is invariant under the adjoint action of
$K_{\mathbb{C}}$ on $\mathfrak{p}_{\mathbb{C}}.$ Since almost every point in
$\mathfrak{p}_{\mathbb{C}}$ can be mapped into $\mathfrak{a}_{\mathbb{C}}$ by
the adjoint action of $K_{\mathbb{C}}$, it should be possible to show directly
that the right-hand side of (\ref{isometry2}) is equal to the right-hand side
of (\ref{isometry1}). Something similar to this is done in the compact group
case in \cite[Thm. 2.3]{FMN2}. However, we will follow a different approach
here using intertwining formulas.

\begin{proof}
(Of Theorem \ref{isometry.thm1}.) For radial functions in the complex case we
have a very special \textquotedblleft intertwining formula\textquotedblright%
\ relating the non-Euclidean Laplacian $\Delta_{G/K}$ for $G/K$ and the
Euclidean Laplacian $\Delta_{\mathfrak{p}}$ for $\mathfrak{p}$. Let us
temporarily identify $\mathfrak{p}$ and $G/K$ by means of the exponential
mapping, so that it makes sense to apply both $\Delta_{G/K}$ and
$\Delta_{\mathfrak{p}}$ to the same function. Then the intertwining formula
states that (for radial functions in the complex case)%
\begin{equation}
\Delta_{G/K}f=\frac{1}{\delta}[\Delta_{\mathfrak{p}}-c](\delta f), \label{id1}%
\end{equation}
where $c$ is the norm-squared of half the sum (with multiplicities) of the
positive roots for $G/K.$ (See Proposition V.5.1 in \cite{He3} and the
calculations in the complex case on p. 484.)

One way to prove the identity (\ref{id1}) is to first verify it for spherical
functions, which are known explicitly in the complex case, and then build up
general radial functions from the spherical functions. A more geometric
approach is to work with the bilinear form associated to the Laplacian,
namely,
\begin{equation}
D(f,g):=\int_{G/K}f(x)\Delta g(x)~dx=-\int_{G/K}\nabla f(x)\cdot
\bigtriangledown g(x)~dx, \label{gr}%
\end{equation}
where $f$ and $g$ are, say, smooth real-valued functions of compact support.
If $f$ and $g$ are radial, then at each point $\bigtriangledown f$ and
$\bigtriangledown g$ will be tangent to the maximal flat, since the tangent
space to a generic $K$-orbit is the orthogonal complement of the tangent space
to the flat. From this, it is not hard to see that the \textit{Euclidean}
gradients of $f$ and $g,$ viewed as functions on $\mathfrak{p}$ by means of
the exponential mapping, coincide with the non-Euclidean gradients.

Thinking of $\bigtriangledown f$ and $\bigtriangledown g$ as Euclidean
gradients, let us multiply and divide in (\ref{gr}) by the Jacobian of the
exponential mapping, thus turning the integral into one over $\mathfrak{p}$
with respect to Lebesgue measure. If we then do a Euclidean integration by
parts on $\mathfrak{p},$ we will get one term involving the Laplacian for
$\mathfrak{p}$ and one term involving derivatives of the Jacobian $\delta^{2}$
of the exponential mapping. With a bit of manipulation, this leads to an
expression of the same form as (\ref{id1}), except with the constant $c$
replaced by the \textit{function} $\Omega:=\Delta_{\mathfrak{p}}%
(\delta)/\delta.$ (See Proposition V.5.1 in \cite{He3} or Theorem II.3.15 in
\cite{He2}.)

Now, up to this point, the argument is valid for an arbitrary symmetric space
of the noncompact type. What is special about the complex case is that in this
case \cite[p. 484]{He3}, we have that $\Delta_{\mathfrak{p}}(\delta)=c\delta,$
so that $\Omega$ is a constant. It turns out that having $\Delta
_{\mathfrak{p}}(\delta)$ be a constant multiple of $\delta$ is equivalent to
having $\Delta_{G/K}(\delta^{-1})$ be a constant multiple (with the opposite
sign) of $\delta^{-1}.$ It is shown in detail in \cite[Sect. 2]{HSt} that this
last condition holds precisely when we have a reduced root system with all
roots of multiplicity 2, that is, precisely in the complex case.

Meanwhile, formally exponentiating (\ref{id1}) would give%
\begin{equation}
e^{t\Delta_{G/K}/2}f=\frac{1}{\delta}e^{-ct/2}e^{t\Delta_{\mathfrak{p}}%
/2}(\delta f). \label{id2}%
\end{equation}
Indeed, (\ref{id2}) holds for all radial functions $f$ in $L^{2}(G/K),$ in
which case $\delta f$ is an Ad-$K$-invariant function in $L^{2}(\mathfrak{p}%
)$. It is not hard to prove that (\ref{id2}) follows from (\ref{id1}), once we
have established that in the Hilbert space of $L^{2}$ radial functions (on
either $G/K$ or $\mathfrak{p}$), the Laplacian is essentially self-adjoint on
$C^{\infty}$ radial functions of compact support. To prove the essential
self-adjointness, we start with the well-known essential self-adjointness of
the Laplacian on $C_{c}^{\infty},$ as an operator on the full $L^{2}$ space.
We then note that the projection onto the radial subspace (again, on either
$G/K$ or $\mathfrak{p}$) commutes with Laplacian and preserves the space of
$C^{\infty}$ of compact support. From this, essential self-adjointness on
$C^{\infty}$ radial functions of compact support follows by elementary
functional analysis.

Let us rewrite (\ref{id2}) as%
\begin{equation}
e^{t\Delta_{\mathfrak{p}}/2}(\delta f)=e^{ct/2}\delta e^{t\Delta_{G/K}/2}f
\label{id2.5}%
\end{equation}
and then apply the \textit{Euclidean} Segal--Bargmann transform for
$\mathfrak{p}$ to the function $\delta f$ in $L^{2}(\mathfrak{p}).$ The
properties of this transform tell us that $e^{t\Delta_{\mathfrak{p}}/2}(\delta
f)$ has an entire analytic continuation to $\mathfrak{p}_{\mathbb{C}}$ and
that%
\begin{equation}
\int_{\mathfrak{p}}\left\vert \delta(X)f(X)\right\vert ^{2}dX=\int
_{\mathfrak{p}_{\mathbb{C}}}\left\vert e^{t\Delta_{\mathfrak{p}}/2}(\delta
f)(X+iY)\right\vert ^{2}\frac{e^{-\left\vert Y\right\vert ^{2}/2}}{(\pi
t)^{d/2}}dX~dY. \label{id3}%
\end{equation}
Equation (\ref{id2.5}) then tells us that $\delta e^{t\Delta_{G/K}/2}f$ also
has an analytic continuation to $\mathfrak{p}_{\mathbb{C}}$ and that%
\begin{equation}
\int_{\mathfrak{p}}\left\vert \delta(X)f(X)\right\vert ^{2}dX=e^{ct}%
\int_{\mathfrak{p}_{\mathbb{C}}}\left\vert \delta(X+iY)(e^{t\Delta_{G/K}%
/2}f)(X+iY)\right\vert ^{2}\frac{e^{-\left\vert Y\right\vert ^{2}/2}}{(\pi
t)^{d/2}}dX~dY. \label{isom.int}%
\end{equation}
Since the function $\delta e^{t\Delta_{G/K}}f$ has a \textit{holomorphic}
extension to $\mathfrak{p}_{\mathbb{C}},$ the function $e^{t\Delta_{G/K}}f$
has a \textit{meromorphic} extension to $\mathfrak{p}_{\mathbb{C}}.$

Let us now undo the identification of $\mathfrak{p}$ with $G/K$ in
(\ref{isom.int}). The functions $f$ and $e^{t\Delta_{G/K}/2}f$ are radial
functions on $G/K.$ To turn these functions into functions on $\mathfrak{p}$
we compose with the exponential mapping. So we now write $f(e^{X})$ on the
left-hand side of (\ref{isom.int}) and $(e^{t\Delta_{G/K}/2}f)(e^{X+iY})$ on
the right-hand side. We then apply (\ref{volume.relation}) to the left-hand
side of (\ref{isom.int}) to obtain%
\[
\int_{G/K}\left\vert f(x)\right\vert ^{2}dx=e^{ct}\int_{\mathfrak{p}%
_{\mathbb{C}}}\left\vert \delta(X+iY)(e^{t\Delta_{G/K}/2}f)(e^{X+iY}%
)\right\vert ^{2}\frac{e^{-\left\vert Y\right\vert ^{2}/2}}{(\pi t)^{d/2}%
}dX~dY.
\]
This establishes the first part of the theorem.

For the second part of the theorem, suppose that $\Phi$ is meromorphic on
$\mathfrak{p}_{\mathbb{C}},$ radial (that is, invariant under the adjoint
action of $K$ on $\mathfrak{p}_{\mathbb{C}}$), and satisfies%
\[
e^{ct}\int_{\mathfrak{p}_{\mathbb{C}}}\left\vert \Phi(X+iY)\right\vert
^{2}\left\vert \delta(X+iY)\right\vert ^{2}\frac{e^{-\left\vert Y\right\vert
^{2}/t}}{(\pi t)^{d/2}}~dY~dX<\infty.
\]
Then the function $\Phi\delta$ is meromorphic on $\mathfrak{p}_{\mathbb{C}}$
and square-integrable with respect to a measure with a strictly positive
density. This, as pointed out in Section \ref{rn.sec}, implies that
$\Phi\delta$ is actually holomorphic on $\mathfrak{p}_{\mathbb{C}}.$ Then by
the surjectivity of the Segal--Bargmann transform for $\mathfrak{p},$ there
exists a unique function $g$ in $L^{2}(\mathfrak{p})$ with $e^{t\Delta
_{\mathfrak{p}}/2}g=\Phi\delta.$ Since the Segal--Bargmann transform commutes
with the action of $K$, $g$ must also be invariant under the adjoint action of
$K.$ If we let $f$ be the unique function on $G/K$ such that
\[
f(e^{X})=\frac{e^{ct/2}g(X)}{\delta(X)},
\]
then $f$ is radial and in $L^{2}(G/K).$ By (\ref{id2}) we have that
$\delta\cdot e^{t\Delta_{G/K}/2}f=\frac{1}{\delta}e^{t\Delta_{\mathfrak{p}}%
/2}(g)=\Phi$ on $\mathfrak{p}.$ This establishes the existence of the function
$f$ in the second part of the theorem. The uniqueness of this $f$ follows from
the injectivity of the operator $e^{t\Delta_{G/K}/2}$ on $L^{2}(G/K).$
\end{proof}

\begin{proof}
(Of Theorem \ref{isometry.thm2}.) The argument is similar to that in the
preceding proof, except that in this case we use an \textquotedblleft
intertwining formula\textquotedblright\ that relates the non-Euclidean
Laplacian on $G/K$ to the Euclidean Laplacian on $\mathfrak{a}.$ This formula
says that (for radial functions $f$ in the complex case) we have%
\begin{equation}
\left.  (\Delta_{G/K}f)\right\vert _{\mathfrak{a}}=\frac{1}{\eta}%
[\Delta_{\mathfrak{a}}-c](\eta f_{\mathfrak{a}}), \label{id4}%
\end{equation}
where $c$ is the same constant as in (\ref{id1}) and where $f_{\mathfrak{a}}$
is the restriction of $f$ to $\mathfrak{a}$. (See \cite[Prop. II.3.10]{He2}.)
An important difference between this formula and (\ref{id1}) above is that the
function $\eta f_{\mathfrak{a}}$ is Weyl-\textit{anti}-invariant, whereas the
function $\delta f$ in (\ref{id1}) is Ad-$K$-\textit{invariant}.
Exponentiating (\ref{id4}) gives that%
\begin{equation}
e^{t\Delta_{G/K}/2}f=\frac{1}{\eta}e^{-ct/2}e^{t\Delta_{\mathfrak{a}}/2}(\eta
f_{\mathfrak{a}}) \label{heatop0}%
\end{equation}
and so%
\begin{equation}
e^{t\Delta_{\mathfrak{a}}/2}(\eta f_{\mathfrak{a}})=e^{ct/2}\eta
e^{t\Delta_{G/K}/2}f. \label{heatop1}%
\end{equation}

From properties of the Segal--Bargmann transform for $\mathfrak{a}$ we then
see that $e^{t\Delta_{\mathfrak{a}}/2}(\eta f_{\mathfrak{a}})$ has a
holomorphic extension to $\mathfrak{a}_{\mathbb{C}}$ and that%
\begin{equation}
\int_{\mathfrak{a}}\left\vert \eta(H)f(H)\right\vert ^{2}dH=\int
_{\mathfrak{a}_{\mathbb{C}}}\left\vert e^{t\Delta_{\mathfrak{a}}/2}(\eta
f_{\mathfrak{a}})(X+iY)\right\vert ^{2}\frac{e^{-\left\vert Y\right\vert
^{2}/2}}{(\pi t)^{r/2}}dX~dY, \label{isometry3}%
\end{equation}
where $r=\dim\mathfrak{a}.$ Using (\ref{heatop1}) then gives%
\[
\int_{\mathfrak{a}}\left\vert \eta(H)f(H)\right\vert ^{2}dH=e^{ct}%
\int_{\mathfrak{a}_{\mathbb{C}}}\left\vert (e^{t\Delta_{G/K}/2}f)(X+iY)\eta
(X+iY)\right\vert ^{2}\frac{e^{-\left\vert Y\right\vert ^{2}/2}}{(\pi
t)^{r/2}}dX~dY.
\]
We now recognize the left-hand side as being---up to an overall constant---the
$L^{2}$ norm of $f$ over $G/K,$ written using (\ref{volume.relation}) and then
generalized polar coordinates for $\mathfrak{p}$ \cite[Thm. I.5.17]{He2}. We
thus obtain the first part of the theorem. The unspecified constant $B$ in
Theorem \ref{isometry.thm2} comes from the constant $c$ in Theorem I.5.17 of
\cite{He2}.

For the second part of the theorem, assume that $\Phi$ is meromorphic,
Weyl-invariant, and satisfies (\ref{finiteness2}). Then, as in the proof of
Theorem \ref{isometry.thm1}, $\Phi\eta$ is holomorphic. In addition, $\Phi
\eta$ is Weyl-\textit{anti}-invariant. There then exists a Weyl-anti-invariant
function $g$ in $L^{2}(\mathfrak{a})$ with $e^{t\Delta_{\mathfrak{a}}/2}%
g=\Phi\eta.$ We now let $f$ be the function on $A:=\exp\mathfrak{a}$
satisfying%
\[
f(e^{X})=\frac{e^{ct/2}g(X)}{\eta(X)}.
\]
Then $f$ is Weyl-invariant on $A$ and has a unique radial extension to $G/K.$
In light of the comments in the preceding paragraph, this extension of $f$ is
square-integrable over $G/K.$ Then (\ref{heatop0}) tells us that
$e^{t\Delta_{G/K}/2}f=\Phi.$
\end{proof}

\section{Inversion formula\label{inversion.sec}}

In this section, we continue to consider symmetric spaces $G/K$ of the complex
type. However, we now consider functions $f$ on $G/K$ that are not necessarily
radial. We let $F=e^{t\Delta_{G/K}/2}f$ and we want to define the
Segal--Bargmann transform as some sort of analytic continuation of $F.$ In the
radial case, we wrote $F$ in exponential coordinates at the basepoint and then
meromorphically extended $F$ from $\mathfrak{p}$ to $\mathfrak{p}_{\mathbb{C}%
}.$ In the nonradial case, this approach is not appropriate, because we no
longer have a distinguished basepoint. Instead we will analytically continue
$F$ to a neighborhood of $G/K$ inside $G_{\mathbb{C}}/K_{\mathbb{C}}.$

For each $x$ in $G/K,$ we have the geometric exponential map $\exp_{x}%
:T_{x}(G/K)\rightarrow G/K.$ It is not hard to show that this can be
analytically continued to a holomorphic map, also denoted $\exp_{x}$, mapping
the complexified tangent space $T_{x}(G/K)_{\mathbb{C}}$ into $G_{\mathbb{C}%
}/K_{\mathbb{C}}.$ We now consider tubes $T^{R}(G/K)$ in the tangent bundle of
$G/K,$%
\[
T^{R}(G/K)=\left\{  (x,Y)\in T(G/K)\left\vert ~\left\vert Y\right\vert
<R\right.  \right\}  .
\]
Then we let $U_{R}$ be the set in $G_{\mathbb{C}}/K_{\mathbb{C}}$ given by%
\[
U_{R}=\left\{  \exp_{x}(iY)\left\vert (x,Y)\in T^{R}(G/K)\right.  \right\}  .
\]
Here, $\exp_{x}(iY)$ refers to the analytic continuation of the exponential
map at $x.$ (In the $\mathbb{R}^{d}$ case, $\exp_{x}(iy)$ would be nothing but
$x+iy.$)

It can be shown that for all sufficiently small $R,$ $U_{R}$ is an open set in
$G_{\mathbb{C}}/K_{\mathbb{C}}$ and the map $(x,Y)\rightarrow\exp_{x}(iY)$ is
a diffeomorphism of $T^{R}(G/K)$ onto $U_{R}.$ The complex structure on
$T^{R}(G/K)$ obtained by identification with $U_{R}$ is the \textquotedblleft
adapted complex structure\textquotedblright\ of \cite{GStenz1,GStenz2,LS,Sz1}.
Furthermore, Kr\"{o}tz and Stanton have shown that for any $f$ in
$L^{2}(G/K),$ the function $F=e^{t\Delta_{G/K}/2}f$ has an analytic
continuation to $U_{R}$, for all sufficiently small $R$ \cite[Thm. 6.1]{KS2}.
(These results actually hold for arbitrary symmetric spaces of the noncompact
type, not necessarily of the complex type.) We think of the analytic
continuation of $F$ to $U_{R}$ as the Segal--Bargmann transform of $f.$

Our goal in this section is to give an inversion formula that recovers $f$
from the analytic continuation of $F.$ In analogy to the $\mathbb{R}^{d}$ case
and the case of compact symmetric spaces, this should be done by integrating
$F$ over the fibers in $U_{R}\cong T^{R}(G/K).$ Something similar to this is
done by Leichtnam, Golse, and Stenzel in \cite{LGS}, in a very general
setting. However, in \cite[Thm. 0.3]{LGS} there is a term involving
integration over the boundary of the tube of radius $R.$ This boundary term
involves $e^{s\Delta/2}f,$ for all $s<t,$ and an integration with respect to
$s.$ This term is undesirable for us because we wish to think of $t$ as fixed.
In the case of compact symmetric spaces, Stenzel \cite{St} showed that the
boundary term in \cite{LGS} could be removed by letting the radius $R$ tend to
infinity, thus leading to the inversion formula described in Section
\ref{compact.sec}.

Now, our results here will not be based on the work of \cite{LGS}.
Nevertheless, \cite{LGS} and \cite{St} suggest that it is not possible to get
an inversion formula of the sort we want by working with one fixed finite $R$;
rather, we need to let $R$ tend to infinity. Unfortunately, (1) the map
$(x,Y)\rightarrow\exp_{x}(iY)$ ceases to be a diffeomorphism of $T^{R}(G/K)$
with $U_{R}$ for large $R,$ and (2) the function $F=e^{t\Delta_{G/K}/2}f$ does
not in general have a holomorphic (or even meromorphic) extension to $U_{R}$
for large $R.$ For noncompact symmetric spaces of the complex type, we will
nevertheless find a way to let $R$ tend to infinity, by means of a
cancellation of singularities. This leads to an inversion formula that is
analogous to what we have in the compact and Euclidean cases. These results
also lead to a natural conjecture of what the isometry formula should be in
this setting, something we hope to address in a future paper.

\subsection{Inversion for radial functions at identity coset}

Suppose that $f$ is a radial function in $L^{2}(G/K).$ Then we may use the
intertwining formula (\ref{id2.5}) and the inversion formula (\ref{rn.inv}) in
Theorem \ref{rn.thm} to obtain the following. As in the previous section, we
let $\delta$ denote the square root of the Jacobian of the exponential mapping
for $G/K$ and we let $c$ denote the norm-squared of half the sum (with
multiplicities) of the positive roots for $G/K.$

\begin{theorem}
\label{radinv.thm}Let $f$ be a sufficiently regular radial function in
$L^{2}(G/K)$ ($G$ complex) and let $F=e^{t\Delta_{G/K}/2}f.$ Then%
\begin{equation}
f(x_{0})=e^{ct/2}\int_{\mathfrak{p}}F(e^{iY})\delta(iY)\frac{e^{-\left\vert
Y\right\vert ^{2}/2t}}{(2\pi t)^{d/2}}~dY, \label{invert1}%
\end{equation}
with absolute convergence of the integral. Here $x_{0}=e^{0}$ is the identity
coset in $G/K.$
\end{theorem}

Specifically, sufficiently regular may be taken to mean that $f$ has $n$
derivatives in $L^{2}(X)$ (with respect to the Riemannian volume measure) for
some some $n>d/2.$ Note that the proof of Theorem \ref{isometry.thm1} shows
that the function $X\rightarrow F(e^{X})\delta(X)$ has an entire analytic
continuation to $\mathfrak{p}_{\mathbb{C}}.$ Thus the expression
$F(e^{iY})\delta(iY)$ is well defined and nonsingular on all of $\mathfrak{p}%
.$

At first glance, it may seem as if this inversion formula is not very useful,
since it applies only to radial functions and then gives only the value of $f$
at the identity coset. Nevertheless, we will see in the next subsection that
this result leads to a much more general inversion formula that applies to
not-necessarily-radial functions at arbitrary points.

Let us think about how this result compares to the inversion formula that
holds for the compact symmetric space $U/K$ that is dual to $G/K$ (where,
since $G/K$ is of the complex type, $U/K$ is isometric to a compact Lie
group). In (\ref{invert1}), the meromorphically continued function $F(e^{iY})$
is being integrated against the signed measure given by%
\begin{equation}
d\sigma_{t}(Y):=e^{ct/2}\delta(iY)\frac{e^{-\left\vert Y\right\vert ^{2}/2t}%
}{(2\pi t)^{d/2}}~dY,\quad Y\in\mathfrak{p}. \label{pseudo.measure}%
\end{equation}
By analogy with the compact case (Theorem \ref{compact.thm} in the special
form of Theorem \ref{group.thm}), we would expect that the (signed) measure
$\sigma_{t}$ should be the heat kernel measure at the identity coset for the
\textit{compact} symmetric space $U/K$ dual to $G/K,$ written in exponential
coordinates. Clearly, this cannot be precisely true, first, because one does
not have global exponential coordinates on the compact symmetric space and,
second, because the density of the measure in (\ref{pseudo.measure}) assumes
negative values, whereas the heat kernel measure is a positive measure.

Nevertheless, the signed measure in (\ref{pseudo.measure}) turns out to be
very closely related to the heat kernel measure for $U/K.$ Specifically, the
\textit{push-forward} of the measure (\ref{pseudo.measure}) under the
exponential mapping for $U/K$ is precisely the heat kernel measure (at the
identity coset) for $U/K.$ Thus (\ref{pseudo.measure}) itself may be thought
of as an \textquotedblleft unwrapped\textquotedblright\ version of the heat
kernel for $U/K,$ where we think of the exponential map as \textquotedblleft
wrapping\textquotedblright\ the tangent space (in a many-to-one way) around
$U/K.$ What is going on is that the heat kernel at a point $x$ in $U/K$ may be
expressed as a sum of contributions from all of the geodesics connecting the
identity coset to $x.$ The quantity in (\ref{pseudo.measure}) is what we
obtain by breaking apart those contributions, thus obtaining a something on
the space of geodesics, that is, on the tangent space at the identity coset.
Although some geodesics make a negative contribution to the heat kernel, the
heat kernel itself (obtained by summing over all geodesics) is positive at
every point.

\begin{theorem}
\label{pushforward.thm}We may identify $\mathfrak{p}$ with the tangent space
at the identity coset to $U/K$ in such a way that the following holds: The
push-forward of the signed measure $\sigma_{t}$ in (\ref{pseudo.measure})
under the exponential mapping for $U/K$ coincides with the heat kernel measure
for $U/K$ at the identity coset.
\end{theorem}

Let us now recall the construction \cite[Sect. V.2]{He1} of $U/K$ and explain
how $\mathfrak{p}$ is identified with the tangent space to $U/K$ at the
identity coset. Let $G_{\mathbb{C}}$ be the unique simply connected Lie group
whose Lie algebra is $\mathfrak{g}_{\mathbb{C}}.$ Let $\tilde{G}$ be the
connected Lie subgroup of $G_{\mathbb{C}}$ whose Lie algebra is $\mathfrak{g}%
.$ For notational simplicity, let us assume that the inclusion of
$\mathfrak{g}$ into $\mathfrak{g}_{\mathbb{C}}$ induces an isomorphism of $G$
with $\tilde{G}.$ (Every symmetric space of the noncompact type can be
realized as $G/K$ with $G$ having this property.) Let $U$ be the connected Lie
subgroup of $G_{\mathbb{C}}$ whose Lie algebra is $\mathfrak{\ u}%
=\mathfrak{k}+i\mathfrak{p}$. Then the connected Lie subgroup of $U$ with Lie
algebra $\mathfrak{k}$ is simply the group $K.$

We consider the quotient manifold $U/K$ and we identify the tangent space at
the identity coset in $U/K$ with $\mathfrak{p}_{\ast}:=i\mathfrak{p}.$ If we
use the multiplication by $i$ map to identify $\mathfrak{p}$ with
$\mathfrak{p}_{\ast},$ then we may transport the inner product on
$\mathfrak{p}$ to $\mathfrak{p}_{\ast}.$ There is then a unique $U$-invariant
Riemannian metric on $U/K$ coinciding with this inner product at the identity
coset. With this Riemannian metric, $U/K$ becomes a simply connected symmetric
space of the compact type, and is called the \textquotedblleft
dual\textquotedblright\ of the symmetric space $G/K$ of the noncompact type.
The duality construction is valid starting with any symmetric space of the
noncompact type, producing a symmetric space of the compact type (and a very
similar procedure goes from compact type to noncompact type). If one begins
with a noncompact symmetric space of the complex type, the dual compact
symmetric space will be isometric to a compact Lie group with a bi-invariant measure.

\begin{proof}
(Of Theorem \ref{radinv.thm}.) Let us again identify $G/K$ with $\mathfrak{p}$
by means of the exponential mapping at the identity coset. Suppose $f$ is a
radial function square-integrable with respect to the Riemannian volume
measure for $G/K.$ Then $\delta f$ is a radial function square-integrable with
respect to the Lebesgue measure for $\mathfrak{p}.$ According to (\ref{id2.5})
in the previous section, we have%
\begin{equation}
e^{t\Delta_{\mathfrak{p}}/2}(\delta f)=e^{ct/2}\delta e^{t\Delta_{G/K}/2}f
\label{identity}%
\end{equation}
If $\delta f$ is \textquotedblleft sufficiently regular,\textquotedblright%
\ then we may apply the inversion formula for the Euclidean Segal--Bargmann
transform ((\ref{rn.inv}) in Theorem \ref{rn.thm}) to the function $\delta f.$
Noting that $\delta(0)=1,$ applying the inversion at the origin gives%
\[
f(0)=(\delta f)(0)=e^{ct/2}\int_{\mathfrak{p}}F(iY)\delta(iY)\frac
{e^{-\left\vert Y\right\vert ^{2}/2t}}{(2\pi t)^{d/2}}~dY,
\]
with absolute convergence of the integral, where $F$ is the meromorphic
extension of $e^{t\Delta_{G/K}/2}f.$ To undo the identification of $G/K$ with
$\mathfrak{p},$ we simply replace $f(0)$ with $f(e^{0})$ and $F(Y)$ with
$F(e^{iY}).$ This establishes Theorem \ref{radinv.thm}, provided that $\delta
f$ is \textquotedblleft sufficiently regular.\textquotedblright

To address the regularity condition, we recall the intertwining formula
(\ref{id1}). From this formula it is not hard to show that if $f$ is radial
and in the domain of $(cI-\Delta_{G/K})^{n/2}$ for some $n,$ then $\delta f$
is in the domain of $(cI-\Delta_{\mathfrak{p}})^{n/2}.$ However, the domain of
$(cI-\Delta_{G/K})^{n/2}$ is precisely the Sobolev space of functions on $G/K$
having $n$ derivatives in $L^{2}.$ Thus if $f$ is in this Sobolev space with
$n>d/2,$ $\delta f$ will be in the corresponding Sobolev space on
$\mathfrak{p}$ and $\delta f$ will indeed be \textquotedblleft sufficiently
regular\textquotedblright\ in the sense of \cite[Sect. 2.1]{range}.
\end{proof}

\begin{proof}
(Of Theorem \ref{pushforward.thm}.) We make use of the formula for the heat
kernel function (at the identity) on a compact Lie group, as originally
obtained by \`{E}skin \cite{E} and rediscovered by Urakawa \cite{U}. We
continue to use symmetric space notation for $U/K$, rather than switching to
group notation. Nevertheless, the following formula is valid \textit{only} in
the case that $U/K$ is isometric to a compact Lie group (which is precisely
when $G/K$ is of the complex type). We think of $\mathfrak{p}_{\ast
}:=i\mathfrak{p}$ as the tangent space to $U/K$ at the identity coset and we
write $e^{Y}$ for the exponential (in the geometric sense) of $Y\in
\mathfrak{p}_{\ast}.$ For any maximal commutative subspace $\mathfrak{a}$ of
$\mathfrak{p},$ the space $\mathfrak{a}_{\ast}:=i\mathfrak{a}$ is a maximal
commutative subspace of $\mathfrak{p}_{\ast}$ (and every maximal commutative
subspace of $\mathfrak{p}_{\ast}$ arises in this way). Given a fixed such
subspace $\mathfrak{a}_{\ast},$ the set $A_{\ast}=\exp(\mathfrak{a}_{\ast})$
is a maximal flat in $U/K$ and $A_{\ast}$ is isometric to a flat Euclidean
torus. Let $\Gamma\subset\mathfrak{a}_{\ast}$ denote the kernel of the
exponential mapping for $\mathfrak{a}_{\ast},$ so that $\Gamma$ is a lattice
in $\mathfrak{a}_{\ast}.$

We now let $\rho_{t}$ denote the fundamental solution at the identity coset to
the heat equation $\partial u/\partial t=\frac{1}{2}\Delta u$ on $U/K$. The
heat kernel formula asserts that for any maximal commutative subspace
$\mathfrak{a}_{\ast}$ of $\mathfrak{p}_{\ast}$ we have%
\begin{equation}
\rho_{t}(e^{H})=\frac{e^{ct/2}}{(2\pi t)^{d/2}}\sum_{\gamma\in\Gamma}%
j^{-1/2}(H+\gamma)e^{-\left\vert H+\gamma\right\vert ^{2}/2t},\quad
H\in\mathfrak{a}_{\ast}. \label{ur.form}%
\end{equation}
The function $\rho_{t}$ is the heat kernel \textit{function}, that is, the
density of the heat kernel \textit{measure} (at the identity coset) with
respect to the (un-normalized) Riemannian volume measure on $U/K.$

In this formula, $d=\dim(U/K)$ and $c$ is the norm squared of half the sum
(with multiplicities) of the positive roots for $U/K.$ Since (it is easily
seen) the roots and multiplicities for $U/K$ are the same as for $G/K,$ this
definition of $c$ agrees with the one made earlier in this section. Meanwhile,
the function $j$ is the Jacobian of the exponential mapping for $U/K$,
$j^{1/2}$ is the unique \textit{smooth} square root of $j$ that is positive
near the origin, and $j^{-1/2}$ is the reciprocal of $j^{1/2}.$ Explicitly,
for $H$ in $\mathfrak{a}_{\ast}$ we have%
\begin{equation}
j^{1/2}(H)=\prod_{\alpha\in R^{+}}\frac{\sin\alpha(H)}{\alpha(H)}%
,\label{j.half}%
\end{equation}
where $R^{+}$ is a set of positive roots for $U/K.$ Note that $j^{1/2}$ takes
on both positive and negative values; the non-negative square root of $j$ is
not a smooth function. Properly, the formula (\ref{ur.form}) is valid only for
$H$ such that $j(H)$ is nonzero, in which case $j(H+\gamma)$ will be nonzero
for all $\gamma\in\Gamma.$ However, since $\rho_{t}$ is continuous, we may
then extend the right-hand side by continuity to all $H\in\mathfrak{a}_{\ast
}.$

Since the roots for $U/K$ are the same as for $G/K$ (under the obvious
identification of $\mathfrak{p}_{\ast}$ with $\mathfrak{p}$), comparing the
formula (\ref{delta.form}) with (\ref{j.half}) gives that
\begin{equation}
j^{1/2}(Y)=\delta(iY) \label{j.delta}%
\end{equation}
for all $Y$ in $\mathfrak{p}\cong\mathfrak{p}_{\ast}.$

The formula (\ref{ur.form}) is not quite what is given in \cite{E} or
\cite{U}, but can be deduced from those papers. Our formula differs from the
one in Urakawa by some factors of 2 having to do with group notation versus
symmetric space notation, some additional factors of 2 having to do with
different normalizations of the heat equation, and an overall constant coming
from different normalizations of the measure on $U/K.$

Now, a \textquotedblleft generic\textquotedblright\ point in $U/K$ (in a sense
to be specified later) is contained in a \textit{unique} maximal flat
$A_{\ast}.$ If $x$ is contained in a unique maximal flat $A_{\ast}$ and if
$e^{Y}=x$ for some $Y$ in $\mathfrak{p}_{\ast},$ then we must have
$Y\in\mathfrak{a}_{\ast}.$ (If $Y$ were not in $\mathfrak{a}_{\ast},$ then $Y$
would be contained in some maximal commutative subspace $\mathfrak{b}_{\ast
}\neq\mathfrak{a}_{\ast}$ and then $x$ would be in the maximal flat $B_{\ast
}\neq A_{\ast}.$) Fix such a point $x$ and pick one $H$ in $\mathfrak{a}%
_{\ast}$ with $e^{H}=x.$ Then the elements of the form $Y=H+\gamma,$ with
$\gamma$ in $\Gamma,$ represent \textit{all} the points in $\mathfrak{p}%
_{\ast}$ with $e^{Y}=x.$ This means that for a generic point $x=e^{H},$ the
sum in (\ref{ur.form}) may be thought of as a sum over all the geodesics
connecting the identity coset to $x.$ If we also make use of (\ref{j.delta}),
we may rewrite (\ref{ur.form}) as%
\begin{equation}
\rho_{t}(x)=\frac{e^{ct/2}}{(2\pi t)^{d/2}}\sum_{\left\{  Y\in\mathfrak{p}%
_{\ast}|e^{Y}=x\right\}  }\delta^{-1}(iY)e^{-\left\vert Y\right\vert ^{2}/2t},
\label{ur.form2}%
\end{equation}
whenever $x$ in $U/K$ is contained in a unique maximal flat.

We are now in a position to understand why Theorem \ref{pushforward.thm}
holds. If we push forward the signed measure in $\sigma_{t}$ in
(\ref{pseudo.measure}), we will get a factor of $1/j(Y)$ ($=1/\delta^{2}(iY)$)
from the change of variables formula, which will change the $\delta$ in
(\ref{pseudo.measure}) to $\delta^{-1}.$ The density of the pushed-forward
measure at a generic point $x$ in $U/K$ will then be a sum over $\left\{
Y|e^{Y}=x\right\}  $ of the density in (\ref{pseudo.measure}) multiplied by
$1/\delta(iY),$ which is precisely what we have in (\ref{ur.form2}). This is
what Theorem \ref{pushforward.thm} asserts.

To make the argument in the preceding paragraphs into a real proof, we need to
attend to a few technicalities, including an appropriate notion of
\textquotedblleft generic.\textquotedblright\ We call an element $Y$ of
$\mathfrak{p}_{\ast}$ \textit{singular} if there exist a maximal commutative
subspace $\mathfrak{a}$ containing $Y,$ a root $\alpha$ for $\mathfrak{a}$,
and an integer $n$ such that $\alpha(Y)=n\pi$; we call $Y$ \textit{regular}
otherwise. We call an element $x$ of $U/K$ singular if $x$ can be expressed as
$x=e^{Y}$ for some singular element $Y\in\mathfrak{p}_{\ast}$; we call $x$
regular otherwise. It can be shown that $e^{Y}$ is regular whenever $Y$ is
regular (this is not immediately evident from the definitions). In both
$\mathfrak{p}_{\ast}$ and $U/K,$ the singular elements form a closed set of
measure zero. Thus in pushing forward the signed measure $\sigma_{t},$ we may
simply ignore the singular points and regard the exponential mapping as taking
the open set of regular elements in $\mathfrak{p}_{\ast}$ onto the open set of
regular elements in $U/K.$ (See Sections VII.2 and VII.5 of \cite{He1}.)

If $x$ is regular and $x=e^{Y},$ then (by definition) $Y$ is regular and it
follows that $j(Y)$ is nonzero. Furthermore, if $x$ is regular then (it can be
shown) $x$ is contained in a unique maximal flat. Thus (\ref{ur.form2}) is
valid for all regular elements. Furthermore, it is easily seen that the
function $j(Y)=\delta(iY)$ has constant sign on each connected component of
the set of regular elements in $\mathfrak{p}_{\ast}.$ Finally, we note that
the exponential mapping is a local diffeomorphism near each regular element of
$\mathfrak{p}_{\ast},$ since the Jacobian of the exponential mapping is
nonzero at regular points. From all of this, it is not hard to use a partition
of unity to show that the argument given above is correct.
\end{proof}

\subsection{Inversion for general functions}

At each point $x$ in $G/K,$ we have the geometric exponential mapping,
$\exp_{x},$ mapping the tangent space $T_{x}(G/K)$ into $G/K.$ We have also
the square root of the Jacobian of the exponential mapping for $\exp_{x},$
denoted $\delta_{x}.$ Now, the action of $G$ gives a linear isometric
identification of $T_{x}(G/K)$ with $T_{x_{0}}(G/K)\cong\mathfrak{p}.$ This
identification is unique up to the adjoint action of $K$ on $\mathfrak{p}.$
Under any such identification, the function $\delta_{x}$ will coincide with
the function $\delta=\delta_{x_{0}}$ considered in the previous section. Thus,
in a slight abuse of notation, we let $\delta$ stand for the square root of
the Jacobian of $\exp_{x}$ at any point $x.$ For example, in the case of
3-dimensional hyperbolic space (with the usual normalization of the metric),
we have $\delta(X)=\sinh\left\vert X\right\vert /\left\vert X\right\vert $
(for all $x$). For any $x,$ the function $\delta$ has an entire analytic
continuation to the complexified tangent space at $x.$

\begin{theorem}
\label{inv.thm2}Let $f$ be in $L^{2}(G/K)$ ($G$ complex) and let
$F=e^{t\Delta_{G/K}/2}f.$ Then define%
\begin{equation}
L(x,R)=e^{ct/2}\int_{\left\vert Y\right\vert \leq R}F(\exp_{x}(iY))\delta
(iY)\frac{e^{-\left\vert Y\right\vert ^{2}/2t}}{(2\pi t)^{d/2}}~dY,
\label{l.def}%
\end{equation}
for all sufficiently small $R.$

Then for each $x,$ $L(x,R)$ admits a real-analytic continuation in $R$ to
$(0,\infty).$ Furthermore, if $f$ is sufficiently regular, then%
\begin{equation}
f(x)=\lim_{R\rightarrow\infty}L(x,R) \label{fake.inv0}%
\end{equation}
for all $x$ in $G/K.$ Thus we may write, informally,%
\begin{equation}
f(x)=\text{ \textquotedblleft}\lim_{R\rightarrow\infty}%
\text{\textquotedblright\ }e^{ct/2}\int_{\left\vert Y\right\vert \leq R}%
F(\exp_{x}iY)\delta(iY)\frac{e^{-\left\vert Y\right\vert ^{2}/2t}}{(2\pi
t)^{d/2}}~dY, \label{fake.inv}%
\end{equation}
with the understanding that the right-hand side is to be interpreted literally
for small $R$ and by analytic continuation in $R$ for large $R.$
\end{theorem}

As in the radial case, \textquotedblleft sufficiently
regular\textquotedblright\ may be interpreted to mean that $f$ has $n$
derivatives in $L^{2}(G/K),$ for some $n$ with $n>d/2.$

The formula (\ref{fake.inv}) should be thought of as the noncompact dual to
the compact group formula (\ref{group.inv}) in Theorem \ref{group.thm}.
Specifically (as in (\ref{j.delta})), $\delta(iY)$ is nothing but the square
root of the Jacobian of the exponential mapping for the dual compact symmetric
space $U/K,$ so that this factor in (\ref{fake.inv}) is dual to the factor of
$j(Y)^{1/2}$ in (\ref{group.inv}). The positive constant $c$ has the same
value in (\ref{fake.inv}) as in (\ref{group.inv}) (because the roots and
multiplicities for $G/K$ and $U/K$ are the same); the change from $e^{-ct/2}$
in (\ref{group.inv}) to $e^{ct/2}$ in (\ref{fake.inv}) is part of the duality.
(For example, the exponential factors are related to the scalar curvature,
which is negative in $G/K$ and positive in $U/K.$)

Let us think about why $L(x,R)$ admits an analytic continuation in $R,$
despite the singularities that develop in $F(\exp_{x}(iY))$ when $Y$ is not
small. The key observation is that the signed measure in the definition of
$L(x,R)$ (denoted $\sigma_{t}$ in (\ref{pseudo.measure})) is radial. Thus the
integral in (\ref{l.def}) only \textquotedblleft sees\textquotedblright\ the
part of $F(\exp_{x}(iY))$ that is radial as a function of $Y.$ Taking the
radial part of $F(\exp_{x}(iY))$ eliminates many of the singularities. The
singularities that remain in the radial part of $F(\exp_{x}(iY))$ are then of
a \textquotedblleft universal\textquotedblright\ nature, coming essentially
from the singularities in the analytically continued spherical functions for
$G/K.$ These remaining singularities are canceled by the zeros in the function
$\delta(iY).$ See Section 5 of the expository paper \cite{range} for further
discussion of the cancellation of singularities.

\begin{proof}
For any $x$ in $G/K,$ we let $K_{x}$ denote the subgroup of $G$ that
stabilizes $x.$ (This group is conjugate in $G$ to $K.$) For any continuous
function $\phi$ on $G/K,$ we let $\phi^{(x)}$ denote the \textquotedblleft
radial part of $\phi$ relative to $x$,\textquotedblright\ given by%
\[
\phi^{(x)}(y)=\int_{K_{x}}\phi(k\cdot y)~dk,
\]
where $dk$ is the normalized Haar measure on $K_{x}.$

We wish to reduce the inversion formula in Theorem \ref{inv.thm2} to the
radial case in Theorem \ref{radinv.thm}. Of course, there is nothing special
about the identity coset in Theorem \ref{radinv.thm}; the same result applies
to functions that are radial with respect to any point $x$ in $G/K.$ Now, note
that%
\[
f^{(x)}(x)=f(x)
\]
and that (since the heat operator commutes with the action of $K_{x}$)%
\[
e^{t\Delta_{G/K}/2}(f^{(x)})=(e^{t\Delta_{G/K}/2}f)^{(x)}=F^{(x)}.
\]
Furthermore, if $f$ is sufficiently regular, then so is $f^{(x)}.$

Thus, by Theorem \ref{radinv.thm} (extended to functions that are radial
around $x$) we have%
\begin{align}
f(x)  &  =f^{(x)}(x)\nonumber\\
&  =\int_{T_{x}(G/K)}e^{t\Delta_{G/K}/2}(f^{(x)})(\exp_{x}(iY))\delta
(iY)\frac{e^{-\left\vert Y\right\vert ^{2}/2t}}{(2\pi t)^{d/2}}~dY\nonumber\\
&  =\int_{T_{x}(G/K)}F^{(x)}(\exp_{x}(iY))\delta(iY)\frac{e^{-\left\vert
Y\right\vert ^{2}/2t}}{(2\pi t)^{d/2}}~dY. \label{fx.inv}%
\end{align}
Note that the function $X\rightarrow F^{(x)}(\exp_{x}(X))\delta(X)$ has an
entire analytic continuation to $T_{x}(G/K)_{\mathbb{C}}$ and therefore
$F^{(x)}(\exp_{x}(iY))\delta(iY)$ is nonsingular for all $Y.$

Now, the action of $K_{x}$ commutes with $\exp_{x}$ and with analytic
continuation from $T_{x}(G/K)$ to $T_{x}(G/K)_{\mathbb{C}}.$ Thus%
\[
F^{(x)}(\exp_{x}(iY))=\int_{K_{x}}F(\exp_{x}(i\mathrm{Ad}_{k}(Y)))~dk.
\]
From this and the fact that $\delta(iY)$ and $\left\vert Y\right\vert ^{2}$
are radial functions of $Y,$ we obtain the following: We may replace
$F(\exp_{x}(iY))$ in (\ref{l.def}) with $F^{(x)}(\exp_{x}(iY))$ without
affecting the value of the integral. This establishes the existence of the
analytic continuation in $R$ of $L(x,R)$: The analytic continuation is given
by%
\[
L(x,R)=e^{ct/2}\int_{\left\vert Y\right\vert \leq R}F^{(x)}(\exp
_{x}(iY))\delta(iY)\frac{e^{-\left\vert Y\right\vert ^{2}/2t}}{(2\pi t)^{d/2}%
}~dY
\]
for all $R.$ (This expression is easily seen to be analytic in $R.$) Letting
$R$ tend to infinity gives the inversion formula (\ref{fake.inv0}), by
(\ref{fx.inv}).
\end{proof}

\section{Review of the compact case\label{compact.sec}}

In order to put our results for noncompact symmetric spaces of the complex
type into perspective, we review here the main results from the compact case.
We describe first the results of Stenzel \cite{St} for general compact
symmetric spaces. Then we describe how those results simplify in the case of a
compact Lie group, recovering results of \cite{H1,H2}. Finally, we describe a
recent result of Florentino, Mour\~{a}o, and Nunes \cite{FMN2} for radial
functions in the compact group case. Our isometry formula for radial functions
in the complex case (especially Theorem \ref{isometry.thm2}) should be
compared to the result of \cite{FMN2}, as described in our Section
\ref{groupradial.sec}. Our inversion formula for general functions (Theorem
\ref{inv.thm2}) should be compared to the inversion formula in the compact
group case, as described in (\ref{group.inv}) of Theorem \ref{group.thm}.

For additional information on the Segal--Bargmann transform for compact groups
and compact symmetric spaces, see the expository papers \cite{bull, range}.
See also \cite{HM1,HM2} for more on the special case of spheres.

We make use here of standard results about compact symmetric spaces (see, for
example, \cite{He1}) as well as results from Section 2 of \cite{St} (or
Section 8 of \cite{LGS}).

\subsection{The general compact case\label{compactgen.sec}}

We consider a compact symmetric space $X$, assumed for simplicity to be simply
connected. Suppose that $U$ is a compact, simply connected Lie group
(necessarily semisimple) and that $\sigma$ is an involution of $U.$ Let $K$ be
the subgroup of $U$ consisting of the elements fixed by $\sigma.$ Then $K$ is
automatically a closed, connected subgroup of $U$. Consider the quotient
manifold $X:=U/K$, together with any Riemannian metric on $U/K$ that is
invariant under the action of $U.$ Then $X$ is a simply connected compact
symmetric space, and every simply connected compact symmetric space arises in
this way. We will assume (without loss of generality) that $U$ acts in a
locally effective way on $X,$ that is, that the set of $u\in U$ for which $u$
acts trivially on $X$ is discrete. Under this assumption, the $U$ and $\sigma$
are unique up to isomorphism for a given $X,$ and $U$ is isomorphic to the
universal cover of the identity component of the isometry group of $X.$

We consider the complexification of the group $U$, denoted $U_{\mathbb{C}}.$
Since we assume $U$ is simply connected, $U_{\mathbb{C}}$ is just the unique
simply connected group whose Lie algebra is $\mathfrak{u}_{\mathbb{C}%
}:=\mathfrak{u}+i\mathfrak{u}$ (where $\mathfrak{u}$ is the Lie algebra of
$U$), and $U$ sits inside $U_{\mathbb{C}}$ as a maximal compact subgroup. We
also let $K_{\mathbb{C}}$ denote the connected Lie subgroup of $U_{\mathbb{C}%
}$ whose Lie algebra is $\mathfrak{k}_{\mathbb{C}}:=\mathfrak{k}%
+i\mathfrak{k}$ (where $\mathfrak{k}$ is the Lie algebra of $K$). Then
$K_{\mathbb{C}}$ is always a closed subgroup of $U_{\mathbb{C}}.$ We may
introduce the \textquotedblleft complexification\textquotedblright\ of $U/K,$
namely, the complex manifold
\[
X_{\mathbb{C}}:=U_{\mathbb{C}}/K_{\mathbb{C}}.
\]
It can be shown that $K_{\mathbb{C}}\cap U=K$; as a result, the inclusion of
$U$ into $U_{\mathbb{C}}$ induces an inclusion of $U/K$ into $U_{\mathbb{C}%
}/K_{\mathbb{C}}$.

We write $g\cdot x$ for the action of an element $g$ in $U_{\mathbb{C}}$ on a
point $x$ in $U_{\mathbb{C}}/K_{\mathbb{C}}$ and we let $x_{0}$ denote the
identity coset in $U/K\subset U_{\mathbb{C}}/K_{\mathbb{C}}.$

\begin{definition}
The \textbf{Segal--Bargmann transform} for $U/K$ is the map
\[
C_{t}:L^{2}(U/K)\rightarrow\mathcal{H}(U_{\mathbb{C}}/K_{\mathbb{C}})
\]
given by%
\[
C_{t}f=\text{ analytic continuation of }e^{t\Delta/2}f.
\]
Here $e^{t\Delta/2}$ is the time-$t$ forward heat operator and the analytic
continuation is from $U/K$ to $U_{\mathbb{C}}/K_{\mathbb{C}}$ with $t$ fixed.
\end{definition}

It follows from \cite[Sect. 4]{H1} (applied to $K$-invariant functions on $U$)
that for any $f$ in $L^{2}(U/K)$ (with respect to the Riemannian volume
measure), $e^{t\Delta/2}f$ has a unique analytic continuation from $U/K$ to
$U_{\mathbb{C}}/K_{\mathbb{C}}.$

At each point $x$ in $U/K,$ we have the geometric exponential map%
\[
\exp_{x}:T_{x}(U/K)\rightarrow U/K.
\]
(If $\gamma$ is the unique geodesic with $\gamma(0)=x$ and $\dot{\gamma
}(0)=Y,$ then $\exp_{x}(Y)=\gamma(1).$) For each $x,$ the map $\exp_{x}$ can
be analytically continued to a holomorphic map of the complexified tangent
space $T_{x}(U/K)_{\mathbb{C}}$ into $U_{\mathbb{C}}/K_{\mathbb{C}}.$

\begin{proposition}
[Identification of $T(X)$ with $X_{\mathbb{C}}$]\label{complex.prop}The map
$\Phi:T(U/K)\rightarrow U_{\mathbb{C}}/K_{\mathbb{C}}$ given by%
\[
\Phi(x,Y)=\exp_{x}(iY),\quad x\in U/K,~Y\in\mathfrak{p}_{x}%
\]
is a diffeomorphism. On right-hand side of the above formula, $\exp_{x}(iY)$
refers to the analytic continuation of geometric exponential map.
\end{proposition}

From the point of view of quantization, we should really identify
$U_{\mathbb{C}}/K_{\mathbb{C}}$ with the \textit{co}tangent bundle $T^{\ast
}(U/K).$ However, since $U/K$ is a Riemannian manifold we naturally and
permanently identify $T^{\ast}(U/K)$ with the tangent bundle $T(U/K).$ In the
$\mathbb{R}^{d}$ case, $\exp_{x}(iy)$ would be nothing but $x+iy.$

The Lie algebra $\mathfrak{u}$ of $U$ decomposes as $\mathfrak{u}%
=\mathfrak{k}+\mathfrak{p},$ where $\mathfrak{p}$ is the $-1$ eigenspace for
the action of the involution $\sigma$ on $\mathfrak{u}.$ For any $x$ in $U/K$
we define%
\begin{align*}
K_{x}  &  =\mathrm{Ad}_{u}(K)\\
\mathfrak{k}_{x}  &  =\mathrm{Ad}_{u}(\mathfrak{k}),\\
\mathfrak{p}_{x}  &  =\mathrm{Ad}_{u}(\mathfrak{p}),
\end{align*}
where $u$ is any element of $U$ such that $u\cdot x_{0}=x.$ We identify
$\mathfrak{p}=\mathfrak{p}_{x_{0}}$ with the tangent space to $U/K$ at $x_{0}%
$; more generally, we identify $\mathfrak{p}_{x}$ with the tangent space at
$x$ to $U/K$. With this identification, we have%
\[
\exp_{x}(Y)=e^{Y}\cdot x,\quad x\in U/K,~Y\in\mathfrak{p}_{x},
\]
where $e^{Y}\in U$ is the exponential of $Y$ in the Lie group sense.

Now, for each $x\in U/K,$ define a subspace $\mathfrak{g}_{x}$ of
$\mathfrak{u}_{\mathbb{C}}$ by%
\[
\mathfrak{g}_{x}=\mathfrak{k}_{x}+i\mathfrak{p}_{x}.
\]
Then $\mathfrak{g}_{x}$ is a Lie subalgebra of $\mathfrak{u}_{\mathbb{C}}.$ We
let $G_{x}$ denote the connected Lie subgroup of $U_{\mathbb{C}}$ whose Lie
algebra is $\mathfrak{g}_{x}.$ Note that $e^{iY}$ belongs to $G_{x}$ for any
$Y$ in $\mathfrak{p}_{x}.$ Thus, the image under $\Phi$ of $T_{x}(U/K)$ is
contained in the $G_{x}$-orbit of $x.$ In fact, $\Phi(T_{x}(U/K))$ is
precisely the $G_{x}$-orbit of $x,$ and the stabilizer in $G_{x}$ of $x$ is
precisely $K_{x}.$ We record this result in the following.

\begin{proposition}
[Identification of the Fibers]For any $x\in U/K,$ the image inside
$U_{\mathbb{C}}/K_{\mathbb{C}}$ of $T_{x}(U/K)\cong\mathfrak{p}_{x}$ under
$\Phi$ is precisely the orbit of $x$ under $G_{x}.$ Thus the image of
$T_{x}(U/K)$ may be identified naturally with $G_{x}/K_{x}.$
\end{proposition}

Now, each $G_{x}$ is conjugate under the action of $U$ to $G:=G_{x_{0}}.$ Thus
each quotient space $G_{x}/K_{x}$ may be identified with $G/K.$ This
identification depends on the choice of an element $u$ of $U$ mapping $x_{0}$
to $x$ and is therefore unique only up to the action of $K$ on $G/K.$ The
space $G/K$, with an appropriately chosen $G$-invariant Riemannian metric, is
the \textit{dual noncompact symmetric space} to $U/K$. Thus we see that the
map $\Phi$ leads naturally to an identification (unique up to the action of
$K$) of each fiber in $T(U/K)$ with the noncompact symmetric space $G/K.$

Another way to think about the appearance of the geometry of $G/K$ in the
problem is from the following result of Leichtnam, Golse, and Stenzel. If we
analytically continue the metric tensor from $U/K$ to $U_{\mathbb{C}%
}/K_{\mathbb{C}}$ and then restrict to the image of $T_{x}(U/K)$ under $\Phi.$
The result is the \textit{negative of} a Riemannian metric and the image of
$T_{x}(U/K)$, with the resulting Riemannian metric, is isometric to $G/K.$
(See \cite[Prop. 1.17 and Thm 8.5]{LGS}.)

On each fiber $T_{x}(U/K)\cong G/K$ we may then introduce the \textit{heat
kernel measure} (at the identity coset). This measure is given by the
Riemannian volume measure for $G/K$ multiplied by the \textit{heat kernel
function}, denoted $\nu_{t}.$ Under the identification of $T_{x}(U/K)$ with
$G/K,$ the Riemannian volume measure on $G/K$ corresponds to Lebesgue measure
on $T_{x}(U/K)$ multiplied by an explicitly computable Jacobian function $j.$
Thus the heat kernel measure on $T_{x}(U/K)$ is the measure $\nu
_{t}(Y)j(Y)~dY,$ where $dY$ denotes Lebesgue measure.

We are now ready to state the main results of Stenzel's paper \cite{St}.

\begin{theorem}
[Stenzel]\label{compact.thm}Let $f$ be in $L^{2}(U/K)$ and let $F=e^{t\Delta
/2}f.$ Then we have the following results.

1. The \textbf{inversion formula}. If $f$ is sufficiently regular we have%
\begin{equation}
f(x)=\int_{T_{x}(U/K)}F(\exp_{x}(iY))\nu_{t}(Y)j(Y)~dY, \label{compact.inv}%
\end{equation}
with absolute convergence of the integral for all $x.$

2. The \textbf{isometry formula}. For all $f$ in $L^{2}(U/K)$ we have%
\begin{equation}
\int_{U/K}\left\vert f(x)\right\vert ^{2}dx=\int_{U/K}\int_{T_{x}%
(U/K)}\left\vert F(\exp_{x}(iY))\right\vert ^{2}\nu_{2t}(2Y)j(2Y)~2^{d}dY~dx,
\label{compact.isom}%
\end{equation}
where $d=\dim(U/K).$

3. The \textbf{surjectivity theorem}. For any holomorphic function $F$ on
$U_{\mathbb{C}}/K_{\mathbb{C}}\cong T(U/K)$ such that the integral on the
right-hand side of (\ref{compact.isom}) is finite, there exists a unique $f$
in $L^{2}(U/K)$ with $F=C_{t}f.$
\end{theorem}

Note that in (\ref{compact.inv}) we have $\nu_{t}(Y)j(Y),$ whereas in
(\ref{compact.isom}) we have $\nu_{2t}(2Y)j(2Y).$ The smoothness assumption on
$f$ in the inversion formula is necessary to guarantee the convergence in the
inversion formula (\ref{compact.inv}). (The optimal smoothness conditions are
not known in general; Stenzel actually assumes that $f$ is $C^{\infty}.$) As
in the $\mathbb{R}^{n}$ case, the inversion formula in (\ref{compact.inv}) is
\textit{not} the one obtained by viewing the heat operator as a unitary map
(as in the isometry formula) and then taking the adjoint.

The special case of Theorem \ref{compact.thm} in which $U/K$ is a compact Lie
group was established in \cite{H1} and \cite{H2}. (The compact group case is
the one in which $U$ is $H\times H$ and $K$ is the diagonal copy of $H$ inside
$H\times H,$ where $H$ is a simply connected compact Lie group.) See also
\cite{HM1, KR2} for an elementary proof of the isometry formula in the case of
$X=S^{d}.$

The proof of the inversion formula hinges on the duality between the compact
symmetric space $U/K$ and noncompact symmetric space $G/K.$ Specifically, for
a holomorphic function $F$ on $U_{\mathbb{C}}/K_{\mathbb{C}}\cong T(U/K)$ we
have that applying the Laplacian for $G_{x}/K_{x}$ in each fiber and then
restricting to the base gives the negative of the result of first restricting
$F$ to the base and then applying the Laplacian for $U/K.$ So, roughly, the
Laplacian in the fiber is the negative of the Laplacian on the base, on
holomorphic functions. (Compare the result in $\mathbb{C}$ that $d^{2}/dy^{2}$
is the negative of $d^{2}/dx^{2}$ when applied to a holomorphic function.) The
argument is then that applying the \textit{forward} heat equation in the
fibers (by integrating against the heat kernel) has the effect of computing
the \textit{backward} heat equation in the base. The proof of the isometry
formula may then be reduced to the inversion formula; in the process of this
reduction, the change from $\nu_{t}(Y)j(Y)$ to $\nu_{2t}(2Y)j(2Y)$ occurs naturally.

\subsection{The compact group case\label{compactgrp.sec}}

Although the Jacobian function $j$ is explicitly computable for any symmetric
space, the heat kernel $\nu_{t}$ is not. Nevertheless, if $X$ is isometric to
a simply connected compact Lie group with a bi-invariant metric, then the dual
noncompact symmetric space is of the complex type and in this case there is an
explicit formula for $\nu_{t}$ due to Gangolli \cite[Prop. 3.2]{Ga}. Expressed
in terms of the heat kernel \textit{measure}, this formula becomes%
\[
\nu_{t}(Y)j(Y)~dY=e^{-ct/2}j(Y)^{1/2}\frac{e^{-\left\vert Y\right\vert
^{2}/2t}}{(2\pi t)^{d/2}}~dY,
\]
where $dY$ is Lebesgue measure on the fiber, $d=\dim(U/K)=\dim(G/K),$ and $c$
is the norm-squared of half the sum of the positive roots for $X$ (thinking of
$X$ as a symmetric space and counting the roots with their multiplicities). In
the expression for the heat kernel \textit{function}, we would have
$j(Y)^{-1/2}$ instead of $j(Y)^{1/2}.$ Thus we obtain the following.

\begin{theorem}
\label{group.thm}In the compact group case, the \textbf{inversion formula}
take the form%
\begin{equation}
f(x)=e^{-ct/2}\int_{T_{x}(U/K)}F(\exp_{x}(iY))j(Y)^{1/2}\frac{e^{-\left\vert
Y\right\vert ^{2}/2t}}{(2\pi t)^{d/2}}~dY \label{group.inv}%
\end{equation}
and the \textbf{isometry formula} takes the form%
\begin{equation}
\int_{U/K}\left\vert f(x)\right\vert ^{2}dx=e^{-ct}\int_{U/K}\int_{T_{x}%
(U/K)}\left\vert F(\exp_{x}(iY))\right\vert ^{2}j(2Y)^{1/2}\frac
{e^{-\left\vert Y\right\vert ^{2}/t}}{(\pi t)^{d/2}}~dY~dx. \label{group.isom}%
\end{equation}

\end{theorem}

As in the general case, (\ref{group.inv}) holds for sufficiently regular $f$
in $L^{2}(U/K)$ and (\ref{group.isom}) holds for all $f$ in $L^{2}(U/K).$

If we specialize further to the case in which $X$ is the unit sphere $S^{3}$
inside $\mathbb{R}^{4}$ (so that $X$ is isometric to the compact group
$\mathrm{SU}(2)$) and put in the explicit expression for $j(Y)$, the inversion
formula becomes%
\begin{equation}
f(x)=e^{-t/2}\int_{T_{x}(S^{3})}F(\exp_{x}(iY))\frac{\sinh\left\vert
Y\right\vert }{\left\vert Y\right\vert }\frac{e^{-\left\vert Y\right\vert
^{2}/2t}}{(2\pi t)^{3/2}}~dY, \label{s3.inv}%
\end{equation}
and this isometry formula becomes%
\begin{equation}
\int_{S^{3}}\left\vert f(x)\right\vert ^{2}~dx=e^{-t}\int_{S^{3}}\int
_{T_{x}(S^{3})}\left\vert F(\exp_{x}(iY))\right\vert ^{2}\frac{\sinh\left\vert
2Y\right\vert }{\left\vert 2Y\right\vert }\frac{e^{-\left\vert Y\right\vert
^{2}/t}}{(\pi t)^{3/2}}~dY. \label{s3.isom}%
\end{equation}

\subsection{Radial functions in the compact group case\label{groupradial.sec}}

In the compact group case, Florentino, Mour\~{a}o, and Nunes have obtained a
special form of the isometry theorem for radial functions. In this case, the
radial functions (in the symmetric space sense) are simply the class functions
on the compact group. Our Theorem \ref{isometry.thm2} is just the noncompact
dual to Theorem 2.2 of \cite{FMN2}. There does not appear to be an analog of
our Theorem \ref{isometry.thm1} in the compact group case, because there the
exponential mapping is not a global diffeomorphism.

We continue to use symmetric space notation rather than switching to compact
group notation. Let $\mathfrak{a}$ be a maximal commutative subspace of
$\mathfrak{p}$ and let $A=\exp_{x_{0}}(\mathfrak{a}).$ Then $A$ is a
\textquotedblleft maximal flat\textquotedblright\ in $X$ and is isometric to a
flat Euclidean torus. Every point $x$ in $U/K$ can be mapped by the left
action of $K$ into $A.$ Thus a radial function is determined by its values on
$A.$

Because $\mathfrak{a}$ is commutative, we can simultaneously identify the
tangent space at every point in $A$ with $\mathfrak{a}.$ We now define the
\textquotedblleft complexification\textquotedblright\ $A_{\mathbb{C}}$ of $A$
to be the image under $\Phi$ of $T(A)\subset T(X),$ where $\Phi$ is the map in
Proposition \ref{complex.prop}. That is to say, we define%
\[
A_{\mathbb{C}}=\left\{  \left.  \exp_{a}(iY)\in X_{\mathbb{C}}\right\vert a\in
A,~Y\in\mathfrak{a}\right\}  .
\]
The restriction of $\Phi$ to $T(A)$ is a diffeomorphism of $T(A)$ with
$A_{\mathbb{C}}.$ (If we identify $X$ with a compact Lie group $H,$ then $A$
is a maximal torus $T$ inside $H$ and $A_{\mathbb{C}}$ is the complexification
of $T$ inside $H_{\mathbb{C}}.$)

It is convenient to multiply the Riemannian volume measures on $X$ and $A$ by
normalizing factors, so that the total volume of each manifold is equal to 1.
If we used instead the un-normalized Riemannian volume measures, there would
be an additional normalization constant in Theorem \ref{compact.radial}, as in
Theorem \ref{isometry.thm2}. We now let $\eta$ be the Weyl denominator
function on $A.$ This is the smooth, real-valued function, unique up to an
overall sign, with the property that%
\[
\int_{X}f(x)~dx=\frac{1}{\left\vert W\right\vert }\int_{A}f(a)\eta(a)^{2}~da,
\]
for all continuous radial functions $f$ on $X.$ Here $\left\vert W\right\vert
$ is the order of the Weyl group for $X,$ and $dx$ and $da$ are the normalized
volume measures on $X$ and $A,$ respectively. The function $\eta$ has an
entire analytic continuation from $A$ to $A_{\mathbb{C}},$ also denoted
$\eta.$

We are now ready to state Theorem 2.2 of \cite{FMN2}, using slightly different notation.

\begin{theorem}
[Florentino, Mour\~{a}o, and Nunes]\label{compact.radial}Suppose $X$ is
isometric to a compact Lie group with a bi-invariant metric. If $f$ is any
radial function in $L^{2}(X),$ let $F$ denote the analytic continuation to
$X_{\mathbb{C}}$ of $e^{t\Delta/2}f.$ Then%
\begin{equation}
\int_{X}\left\vert f(x)\right\vert ^{2}~dx=\frac{e^{-ct}}{\left\vert
W\right\vert }\int_{A}\int_{\mathfrak{a}}\left\vert F(\exp_{a}(iY)\right\vert
^{2}\left\vert \eta(\exp_{a}(iY)\right\vert ^{2}\frac{e^{-\left\vert
Y\right\vert ^{2}/t}}{(\pi t)^{r/2}}~dY~da, \label{compact.radialisom}%
\end{equation}
Here $r$ is the dimension of $\mathfrak{a},$ the constant $c$ is the same as
in (\ref{group.inv}) and (\ref{group.isom}), $|W|$ is the order of the Weyl
group, and $dx$ and $da$ are the normalized Riemannian volume measures on $X$
and $A,$ respectively.

Furthermore, if $F$ is any Weyl-invariant holomorphic function on
$A_{\mathbb{C}}$ for which the integral on the right-hand side of
(\ref{compact.radialisom}) is finite, then there exists a unique radial
function $f$ in $L^{2}(X)$ such that $F=e^{t\Delta/2}f$ on $A.$
\end{theorem}

Consider, for example, the case in which $X$ is the unit sphere $S^{3}$ in
$\mathbb{R}^{4}$, in which case $X_{\mathbb{C}}$ is the complexified sphere
\[
S_{\mathbb{C}}^{3}:=\left\{  \left.  z\in\mathbb{C}^{4}\right\vert z_{1}%
^{2}+z_{2}^{2}+z_{3}^{2}+z_{4}^{2}=1\right\}  .
\]
Fix the basepoint $x_{0}:=(0,0,0,1).$ In that case, a \textquotedblleft
radial\textquotedblright\ function on $S^{3}$ is one that is invariant under
the rotations that fix $x_{0}.$ If we take $\mathfrak{a}$ to be the
one-dimensional subspace of $T_{x_{0}}(S^{3})$ spanned by the vector
$e_{2}=(0,1,0,0)$, then $A$ is the set%
\begin{equation}
A=\left\{  \left.  (\cos\theta,\sin\theta,0,0)\right\vert \theta\in
\mathbb{R}\right\}  \label{a.form}%
\end{equation}
and $A_{\mathbb{C}}$ is the set of points in $S_{\mathbb{C}}^{3}$ of the same
form as in (\ref{a.form}), except with $\theta$ in $\mathbb{C}.$ In the
$S^{3}$ case, $\left\vert W\right\vert =2,$ $c=1,$ the Weyl denominator is
$2\sin\theta,$ and the normalized measure on $A$ is $d\theta/2\pi.$ Thus
(\ref{compact.radialisom}) becomes%
\begin{align}
&  \int_{S^{3}}\left\vert f(x)\right\vert ^{2}~dx\nonumber\\
&  =\frac{e^{-t}}{2}\int_{0}^{2\pi}\int_{-\infty}^{\infty}\left\vert F\left[
(\cos(\theta+iy),\sin(\theta+iy),0,0)\right]  \right\vert ^{2}~\left\vert
2\sin(\theta+iy)\right\vert ^{2}\frac{e^{-y^{2}/t}}{(\pi t)^{1/2}}%
~dy~\frac{d\theta}{2\pi}.
\end{align}

\section{Concluding remarks\label{conclude.sec}}

In this paper we have established an isometry formula (in two different
versions) for the Segal--Bargmann transform of radial functions and an
inversion formula for the Segal--Bargmann transform of general functions, both
in the case of a noncompact symmetric space of the complex type. Both the
isometry formula and the inversion formula require a cancellation of
singularities, but otherwise they closely parallel the results from the
compact group case. Specifically, Theorem \ref{isometry.thm2} in the complex
case is very similar to Theorem \ref{compact.radial} in the compact group case
and Theorem \ref{inv.thm2} in the complex case is very similar to the
inversion formula in Theorem \ref{group.thm} in the compact group case.
Besides the cancellation of singularities, the main difference between the
formulas in the two cases is the interchange of hyperbolic sine with ordinary
sine. It is natural, then, to look ahead and consider the prospects for
obtaining results in the noncompact setting paralleling \textit{all} of the
results we have for compact symmetric spaces. This would entail extending the
isometry result to nonradial functions and then extending both the isometry
and the inversion results to other noncompact symmetric spaces, beyond those
of the complex type. In \cite{H2} in the compact group case and in \cite{St}
in the general compact symmetric space case, the inversion formula is proved
first and the isometry formula obtained from it. As a result, we fully expect
that the inversion formula we prove here will lead to an isometry formula for
not-necessarily-radial functions in the complex case. A precise statement of
the result we have in mind is given in \cite{range} in the case of hyperbolic 3-space.

Meanwhile, we have recently received a preprint by Kr\"{o}tz, \'{O}lafsson,
and Stanton \cite{KOS} that establishes an isometry formula for general
functions (not necessarily radial) on general symmetric spaces of the
noncompact type (not necessarily of the complex type). However, this isometry
formula does not, at least on the surface, seem parallel to the compact case.
In particular, in the complex case, this isometry formula does \textit{not}
reduce to the one we have in mind, at least not without some substantial
manipulation of the formula in \cite[Thm. 3.3]{KOS}. Nevertheless, the result
of \cite{KOS} is a big step toward understanding the situation for general
symmetric spaces of the noncompact type. There may well be a connection, in
the complex case, between the results of \cite{KOS}\ and the isometry formula
we have in mind, but this remains to be worked out. If the isometry formula
can be understood better for general noncompact symmetric spaces, this
understanding may pave the way for progress on the inversion formula as well.

Note that in the case of compact symmetric spaces, the results take on a
particularly simple and explicit form in the compact group case. (Compare
Theorem \ref{compact.thm} to Theorem \ref{group.thm}.) Our results in this
paper are for the noncompact symmetric spaces of the complex type; this case
is just the dual of the compact group case. Thus, one cannot expect the same
level of explicitness for noncompact symmetric spaces that are not of the
complex type. Instead, we may hope for results that involve some suitably
\textquotedblleft unwrapped\textquotedblright\ version of the heat kernel
measure on the dual compact symmetric space, where in general there will not
be an explicit formula for this unwrapped heat kernel.


\begin{thebibliography}{99999}                                                                                            %


\bibitem[A]{A}A. Ashtekar, J. Lewandowski, D. Marolf, J. Mour\~{a}o, and T.
Thiemann, Coherent state transforms for spaces of connections, \textit{J.
Funct. Anal.} \textbf{135} (1996), 519--551.

\bibitem[Ba]{Ba1}V. Bargmann, On a Hilbert space of analytic functions and an
associated integral transform, \textit{Comm. Pure Appl. Math.} \textbf{14}
(1961), 187--214.

\bibitem[Das1]{Das1}A. Dasgupta, Coherent states for black holes, \textit{J.
Cosmology Astroparticle Phys.} \textbf{08} (2003), 004.

\bibitem[Das2]{Das2}A. Dasgupta, Counting the apparent horizon, preprint.

http://arxiv.org/abs/hep-th/0310069.

\bibitem[DOZ1]{DOZ1}M. Davidson, G. \'{O}lafsson, and G. Zhang, Laguerre
polynomials, restriction principle, and holomorphic representations of
$\mathrm{SL}(2,\mathbb{R})$, \textit{Acta Appl. Math.} \textbf{71} (2002), 261--277.

\bibitem[DOZ2]{DOZ2}M. Davidson, G. \'{O}lafsson, and G. Zhang, Laplace and
Segal-Bargmann transforms on Hermitian symmetric spaces and orthogonal
polynomials, \textit{J. Funct. Anal.} \textbf{204} (2003), 157--195.

\bibitem[DH]{DH1}B. K. Driver and B. C. Hall, Yang-Mills theory and the
Segal-Bargmann transform, \textit{Comm. Math. Phys.} \textbf{201} (1999), 249--290.

\bibitem[E]{E}L. D. \`{E}skin, Heat equation on Lie groups. (Russian) In: In
Memoriam: N. G. Chebotarev, 113--132, Izdat. Kazan. Univ., Kazan, Russia, 1964.

\bibitem[Fo]{Fo}G. B. Folland, \textquotedblleft Harmonic analysis in phase
space.\textquotedblright\ Annals of Mathematics Studies, 122. Princeton
University Press, Princeton, NJ, 1989.

\bibitem[FMN1]{FMN1}C. A. Florentino, J. M. Mour\~{a}o, and J. P. Nunes,
Coherent state transforms and abelian varieties, \textit{J. Funct. Anal.}
\textbf{192} (2002), 410--424.

\bibitem[FMN2]{FMN2}C. A. Florentino, J. M. Mour\~{a}o, and J. P. Nunes,
Coherent state transforms and vector bundles on elliptic curves, \textit{J.
Funct. Anal.} \textbf{204} (2003), 355--398.

\bibitem[FMMN1]{FMMN}C. A. Florentino, P. Matias, J. M. Mour\~{a}o, and J. P.
Nunes, Geometric quantization, complex structures, and the coherent state
transform, \textit{J. Funct. Anal.}, to appear.

http://arxiv.org/abs/math.DG/0402313

\bibitem[FMMN2]{FMMN2}C. A. Florentino, P. Matias, J. M. Mour\~{a}o, and J. P.
Nunes, On the BKS pairing for K\"{a}hler quantizations of the cotangent bundle
of a Lie group, preprint. 

http://arxiv.org/abs/math.DG/0411334

\bibitem[Ga]{Ga}R. Gangolli, Asymptotic behavior of spectra of compact
quotients of certain symmetric spaces, \textit{Acta Math.} \textbf{121}
(1968), 151--192.

\bibitem[GH]{GH}P. Griffiths and J. Harris, Principles of algebraic geometry.
Reprint of the 1978 original. Wiley Classics Library. John Wiley \& Sons,
Inc., New York, 1994.

\bibitem[Gr]{Gr}L. Gross, Uniqueness of ground states for Schr\"{o}dinger
operators over loop groups, \textit{J. Funct. Anal.} \textbf{112} (1993), 373--441.

\bibitem[GM]{GM}L. Gross and P. Malliavin, Hall's transform and the
Segal-Bargmann map. In: \textquotedblleft It\^{o}'s stochastic calculus and
probability theory,\textquotedblright\ 73--116, Springer, Tokyo, 1996.

\bibitem[GS1]{GStenz1}V. Guillemin and M. B. Stenzel, Grauert tubes and the
homogeneous Monge-Amp\`{e}re equation, \textit{J. Differential Geom.}
\textbf{34} (1991), 561--570.

\bibitem[GS2]{GStenz2}V. Guillemin and M. B. Stenzel, Grauert tubes and the
homogeneous Monge-Amp\`{e}re equation. II, \textit{J. Differential Geom.}
\textbf{35} (1992), 627--641.

\bibitem[H1]{H1}B. C. Hall, The Segal-Bargmann \textquotedblleft coherent
state\textquotedblright\ transform for compact Lie groups, \textit{J. Funct.
Anal.} \textbf{122} (1994), 103--151.

\bibitem[H2]{H2}B. C. Hall, The inverse Segal-Bargmann transform for compact
Lie groups, \textit{J. Funct. Anal.} \textbf{143} (1997), 98--116.

\bibitem[H3]{newform}B. C. Hall, A new form of the Segal-Bargmann transform
for Lie groups of compact type, \textit{Canad. J. Math.} \textbf{51} (1999), 816--834.

\bibitem[H4]{mexnotes}B. C. Hall, Holomorphic methods in analysis and
mathematical physics. In: First Summer School in Analysis and Mathematical
Physics (S. P\'{e}rez-Esteva and C. Villegas-Blas, Eds.), 1--59, Contemp.
Math., 260, Amer. Math. Soc., Providence, RI, 2000.

\bibitem[H5]{ymcoherent}B. C. Hall, Coherent states and the quantization of
(1+1)-dimensional Yang-Mills theory, \textit{Rev. Math. Phys.} \textbf{13}
(2001), 1281--1305.

\bibitem[H6]{bull}B. C. Hall, Harmonic analysis with respect to heat kernel
measure, \textit{Bull. Amer. Math. Soc. (N.S.)} \textbf{38} (2001), 43--78.

\bibitem[H7]{geoquant}B. C. Hall, Geometric quantization and the generalized
Segal--Bargmann transform for Lie groups of compact type, \textit{Comm. Math.
Phys.} \textbf{226} (2002), 233-268.

\bibitem[H8]{ergodic}B. C. Hall, The Segal-Bargmann transform and the Gross
ergodicity theorem. In: Finite and infinite dimensional analysis in honor of
Leonard Gross (H.-H. Kuo and A. N. Sengupta, Eds.), 99--116, Contemp. Math.,
317, Amer. Math. Soc., Providence, RI, 2003.

\bibitem[H9]{range}B. C. Hall, The range of the heat operator, preprint.

http://arxiv.org/abs/math.DG/0409308

\bibitem[HL]{HL}B. C. Hall and W. Lewkeeratiyutkul, Holomorphic Sobolev spaces
and the generalized Segal--Bargmann transform, \textit{J. Funct. Anal.},
\textbf{217} (2004) 192--220.

\bibitem[HM1]{HM1}B. C. Hall and J. J. Mitchell, Coherent states on spheres,
\textit{J. Math. Phys.} \textbf{43} (2002), 1211--1236.

\bibitem[HM2]{HM2}B. C. Hall and J. J. Mitchell, The large radius limit for
coherent states on spheres. In: Mathematical results in quantum mechanics
(Taxco, 2001), 155--162, Contemp. Math., 307, Amer. Math. Soc., Providence,
RI, 2002.

\bibitem[HS]{HS}B. C. Hall and A. N. Sengupta, The Segal-Bargmann transform
for path-groups, \textit{J. Funct. Anal.} \textbf{152} (1998), 220--254.

\bibitem[HSt]{HSt}B. C. Hall and M. B. Stenzel, Sharp bounds for the heat
kernel on certain symmetric spaces of non-compact type. In: Finite and
infinite dimensional analysis in honor of Leonard Gross (H.-H. Kuo and A. N.
Sengupta, Eds.), 117-135, Contemp. Math., 317, Amer. Math. Soc., Providence,
RI, 2003.

\bibitem[He1]{He1}S. Helgason, \textquotedblleft Differential geometry, Lie
groups, and symmetric spaces.\textquotedblright\ Corrected reprint of the 1978
original. Graduate Studies in Mathematics, 34. American Mathematical Society,
Providence, RI, 2001.

\bibitem[He2]{He2}S. Helgason, \textquotedblleft Groups and geometric
analysis. Integral geometry, invariant differential operators, and spherical
functions.\textquotedblright\ Corrected reprint of the 1984 original.
Mathematical Surveys and Monographs, 83. American Mathematical Society,
Providence, RI, 2000.

\bibitem[He3]{He3}S. Helgason, \textquotedblleft Geometric analysis on
symmetric spaces.\textquotedblright\ Mathematical Surveys and Monographs, 39.
American Mathematical Society, Providence, RI, 1994.

\bibitem[KR1]{KR1}K. Kowalski and J. Rembieli\'{n}ski, Quantum mechanics on a
sphere and coherent states, \textit{J. Phys. A} \textbf{33} (2000), 6035--6048.

\bibitem[KR2]{KR2}K. Kowalski and J. Rembieli\'{n}ski, The Bargmann
representation for the quantum mechanics on a sphere, \textit{J. Math. Phys.}
\textbf{42} (2001), 4138--4147.

\bibitem[KOS]{KOS}B. Kr\"{o}tz, G. \'{O}lafsson, and R. Stanton, The image of
the heat kernel transform on Riemannian symmetric spaces of the non-compact
type, preprint. 

http://arxiv.org/abs/math.CA/0407391

\bibitem[KS1]{KS1}B. Kr\"{o}tz and R. J. Stanton, Holomorpic extensions of
representations: (I) automorphic functions, \textit{Ann. Math.} \textbf{159}
(2004), 641-724.

\bibitem[KS2]{KS2}B. Kr\"{o}tz and R. J. Stanton, Holomorphic extension of
representations: (II) geometry and harmonic analysis, preprint.

\bibitem[KTX]{KTX}B. Kr\"{o}tz, S. Thangavelu, and Y. Xu, The heat kernel
transform for the Heisenberg group, preprint. 

http://arxiv.org/abs/math.CA/0401243.

\bibitem[Ku]{Ku}I. Kubo, A direct setting of white noise analysis, In:
\textquotedblleft Stochastic Analysis on Infinite Dimensional
Spaces\textquotedblright\ (H. Kunita and H.-H. Kuo, Eds.), Pitman Research
Notes in Mathematics, Vol. 310, Longman House, Essex, England, 1994.

\bibitem[LGS]{LGS}\'{E}. Leichtnam, F. Golse, and M. B. Stenzel, Intrinsic
microlocal analysis and inversion formulae for the heat equation on compact
real-analytic Riemannian manifolds, \textit{Ann. Sci. \'{E}cole Norm. Sup.}
(4) \textbf{29} (1996), 669--736.

\bibitem[LS]{LS}L. Lempert and R. Sz\H{o}ke, Global solutions of the
homogeneous complex Monge-Amp\`{e}re equation and complex structures on the
tangent bundle of Riemannian manifolds, \textit{Math. Ann.} \textbf{290}
(1991), 689--712.

\bibitem[OO]{OO}G. \'{O}lafsson and B. \O rsted, Generalizations of the
Bargmann transform. In: Lie theory and its applications in physics (Clausthal,
1995), 3--14, World Sci. Publishing, River Edge, NJ, 1996

\bibitem[Se1]{Se0}I. E. Segal, Tensor algebras over Hilbert spaces, I,
\textit{Trans. Amer. Math. Soc.} \textbf{81} (1956), 106-134.

\bibitem[Se2]{Se1}I. E. Segal, Mathematical problems of relativistic physics.
In: \textquotedblleft Proceedings of the Summer Seminar, Boulder, Colorado,
1960\textquotedblright\ (M. Kac, Ed.), American Mathematical Society,
Providence, RI, 1963.

\bibitem[Se3]{Se2}I. E. Segal, Mathematical characterization of the physical
vacuum for a linear Bose-Einstein field. (Foundations of the dynamics of
infinite systems. III) \textit{Illinois J. Math.} \textbf{6} (1962), 500--523.

\bibitem[Se4]{Se3}I. E. Segal, The complex-wave representation of the free
Boson field. In: \textquotedblleft Topics in Functional
Analysis\textquotedblright\ (I. Gohberg and M. Kac, Eds.), Advances in
Mathematics Supplementary Studies, Vol. 3, Academic Press, New York, 1978.

\bibitem[St]{St}M. B. Stenzel, The Segal-Bargmann transform on a symmetric
space of compact type, \textit{J. Funct. Anal.} \textbf{165} (1999), 44--58.

\bibitem[Sz1]{Sz1}R. Sz\H{o}ke, Complex structures on tangent bundles of
Riemannian manifolds, \textit{Math. Ann.} \textbf{291} (1991), 409--428.

\bibitem[Sz2]{Sz2}R. Sz\H{o}ke, Adapted complex structures and geometric
quantization, \textit{Nagoya Math. J.} \textbf{154} (1999), 171--183.

\bibitem[Th]{Th}T. Thiemann, Gauge field theory coherent states (GCS) I,
General properties, \textit{Classical Quantum Gravity} \textbf{18} (2001), 2025--2064.

\bibitem[TW1]{TW1}T. Thiemann and O. Winkler, Gauge field theory coherent
states (GCS) II, Peakedness properties, \textit{Classical Quantum Gravity}
\textbf{18} (2001), 2561--2636.

\bibitem[TW2]{TW2}T. Thiemann and O. Winkler, Gauge field theory coherent
states (GCS) III, Ehrenfest theorems, \textit{Classical Quantum Gravity}
\textbf{18} (2001), 4629--4681.

\bibitem[Ty]{Ty}A. Tyurin, \textquotedblleft Quantization, classical and
quantum field theory and theta functions.\textquotedblright\ With a foreword
by Alexei Kokotov. CRM Monograph Series, 21. American Mathematical Society,
Providence, RI, 2003.

\bibitem[U]{U}H. Urakawa, The heat equation on compact Lie group,
\textit{Osaka J. Math.} \textbf{12} (1975), 285--297.

\bibitem[Wo]{Wo}N. M. J. Woodhouse, \textquotedblleft Geometric
quantization.\textquotedblright\ Second edition. Oxford Mathematical
Monographs. Oxford Science Publications. The Clarendon Press, Oxford
University Press, New York, 1992.

\bibitem[Wr]{Wr}K. K. Wren, Constrained quantisation and $\theta$-angles. II,
\textit{Nuclear Phys. B.} \textbf{521} (1998), 471--502.
\end{thebibliography}
\end{document}